\documentclass[prb,notitlepage]{revtex4-1} 

\usepackage{amsmath,bm,bbm,amssymb}    
\usepackage{dsfont}
\usepackage[normalem]{ulem}
\usepackage{graphicx}   
\usepackage{pstricks}

\newcommand{\be}{\begin{equation}}
\newcommand{\ee}{\end{equation}}

\newcommand{\chiinf}{\chi_{_\infty}}
\newcommand{\AD}[1]{\mathcal{A}(\mathcal{D}_{#1})}
\newcommand{\OD}{\overline{\mathcal{D}}}
\newcommand{\DD}{\mathcal{D}}
\usepackage{soul}
\begin{document}

\title{Thinking outside the box: fluctuations and finite size effects}

\author{Dario Villamaina}
\affiliation{Universit\'e Paris-Sud, LPTMS, UMR 8626, Orsay Cedex, F-91405 and
CNRS, Orsay, F-91405}

\author{Emmanuel Trizac}
\affiliation{Universit\'e Paris-Sud, LPTMS, UMR 8626, Orsay Cedex, F-91405 and
CNRS, Orsay, F-91405}

\date{\today}

\begin{abstract}
  The isothermal compressibility of an interacting or non interacting system may
  be extracted from the fluctuations of the number of particles in a
  well chosen control volume. Finite size effects  are
  prevalent and should then be accounted for to obtain a meaningful,
  thermodynamic compressibility. In the traditional computational
  setup where a given simulation box is replicated with periodic
  boundary conditions, we study particle number fluctuations outside
  the box (i.e. when the control volume exceeds the box itself), which
  bear relevant thermodynamic information. We also investigate the related
  problem of extracting the compressibility from the structure factor
  in the small wave-vector limit ($k\to 0$). The calculation
  should be restricted to the discrete set of wave-vectors $\bm k$
  that are compatible with the periodicity of the system, and we assess
  the consequences of considering other
  $\bm k$ values, a widespread error among beginners.
 
\end{abstract}

\maketitle

\section{Context and motivation}

In 1914, Ornstein and Zernike~\cite{OZ14} showed that in an equilibrium 
system at temperature $T$ and density $\rho$,
\be
\frac{\chiinf}{\chiinf^{ig}}  
\, = \,
\frac{\langle N_{\mathcal{D}}^2 \rangle - \langle N_\mathcal{D} \rangle^2}{\langle N_\mathcal{D} \rangle} 
\,=\, 
1 + \rho \int_\mathcal{D} \left[ g(\bm{r})-1 
\right] \, d \bm{r} .
\label{eq:comp}
\ee This emblematic achievement of statistical mechanics --also known
as the compressibility equation-- connects the fluctuations of the
number of particles $N_\mathcal{D}$ in a domain $\mathcal{D}$ of a given infinite system,
to macroscopic isothermal 
compressibility $\chiinf$ 
and to local structure through
the pair correlation function $g(\bm{r})$,~\cite{C87,HM06} 
which encodes the local structure, see Appendix~\ref{app:A}.
Here, $\chiinf^{ig}= (\rho k_B T)^{-1}$ is for the ideal gas compressibility while 
$k_B$ is Boltzmann constant. Equation (\ref{eq:comp}) shows that at a critical point where 
$\chiinf \to \infty$, the
amplitude of fluctuations also diverges and long range order sets
in. Away from the critical point or the phase coexistence regime,
Eq.~(\ref{eq:comp}) provides an operational way to compute the
compressibility in a numerical simulation, but particular attention
should be paid to finite size effects, see~\cite{RWGV08} for a review
and references therein. Indeed, the bulk compressibility follows
from Eq. (\ref{eq:comp}) under grand canonical conditions, that is
when $\mathcal{D}$ is a sub-part of an otherwise infinite (or large enough)
system, such that in addition $\mathcal{D}$ is bigger than the relevant
microscopic length (particle size, correlation length etc). 
In practice however, some simulation techniques like molecular
dynamics operate under microcanonical conditions: fixed available volume,
total number of particles $N$ and energy, with periodic boundary
conditions to emulate bulk properties.~\cite{HM06,AT89} Making use of the 
left hand side equality of (\ref{eq:comp}) to infer $\chiinf$,
one may be tempted to use large volumes $\mathcal D$, in which case
important finite size effects should be accounted for. They are the
central issue under scrutiny here.

The question is the following: considering that in
general the left hand side of Eq. (\ref{eq:comp}) defines a finite
size compressibility $\chi_{_L}(\mathcal{D})$ from the fluctuations of the
number of particles in a volume $\mathcal{D}$ itself enclosed in a cubic
simulation box of size $L$, i.e. 
\begin{equation}
\frac{\chi_{_L}(\mathcal{D})}{\chiinf^{ig}} 
\,\equiv\, 
\frac{\langle N_{\mathcal{D}}^2 \rangle - \langle N_\mathcal{D} \rangle^2}{\langle N_\mathcal{D} \rangle} ,
\label{eq:chiL}
\end{equation}
what is the connection between
$\chi_{_L}(\mathcal{D})$ and the thermodynamic compressibility
$\chiinf$?
According to (\ref{eq:comp}), both quantities should coincide 
when $L$ becomes very large, in which case $\chi_{_L}(\mathcal{D})$
no longer depend on $\mathcal{D}$, 
provided $\mathcal{D}$ it is not too small and does not interfere with microscopic
lengths. How does $\chi_{_L}(\mathcal{D})$ depend on $\mathcal{D}$, when 
the latter volume is not negligible compared to the one available?
We will address below this question in
the presence of periodic boundary conditions. It has already been answered 
in the literature when $\mathcal{D}$ is a sub-volume of the simulation box (see~\cite{SPANISH_AJP}
for a pedagogical account), 
but not in the ``reverse'' perspective, with the simulation box as a sub-part of $\mathcal{D}$ 
[i.e. $L^d < \mathcal{A}(\mathcal{D})$ where $d$ is space dimension, and $\mathcal{A}(\mathcal{D})$ the volume of domain 
$\mathcal{D}$]. The present study of fluctuations
``outside the box'' is, to the best of our knowledge, original. It
offers the possibility to infer thermodynamic information from a
measure that is endowed with strong explicit finite size effects, and
can be used as a teaching material for an advanced undergraduate course in
statistical physics or computational techniques. The discussion is
entirely based on elementary considerations:  basic probability concepts 
like the law of total variance,
and  tools like correlation functions in direct or reciprocal space.
Interesting relations can be derived, that are usually not found in
textbooks.  

The paper is organized as follows. Before addressing in detail
fluctuations ``outside the box'', it is informative to decipher
explicit size effects ``inside the box'', which can be achieved
from appropriate applications of (\ref{eq:comp}). This is worked out
in section \ref{sec:inside}, where known results are recovered,
but from an original angle. We then turn our attention 
outside the box in section \ref{sec:outside}, which requires
more subtle arguments, due to periodicity effects which create
correlations between the particles inside the box, and their images
outside. 
Sections \ref{sec:inside} and \ref{sec:outside}
both are real space studies, and for completeness, we investigate
the structure factor $S_L(\bm{k})$, that can be viewed as a scale dependent 
compressibility living in Fourier space. It is well known that $S_L(\bm{k})$
yields at small $k=|\bm{k}|$ the thermodynamic compressibility, provided
it is computed on the discrete set of ``allowed'' Fourier modes compatible 
with the periodic boundaries.~\cite{HM06} 
As an echo to the out-of-the-box
viewpoint in real space, we analyze in section \ref{sec:structure} 
the consequences of computing $S_L(\bm{k})$ for $k<2\pi/L$,
namely for wavelengths that exceed the box size, and more generally for $\bm{k}$ values that are not within the allowed set. 
While such a procedure clearly is erroneous, it is instructive
to discuss the consequences of such a mistake that is made by many beginners,
and sometimes found in the literature. The more technical aspects of the discussion are 
relegated to appendices.
Our conclusions are finally drawn in section \ref{sec:concl}.  

\section{Finite size effects within the box}
\label{sec:inside}

We begin our discussion by emphasizing that 
in Eq.~\eqref{eq:comp}, the left hand side equality
\begin{equation}
\frac{\chiinf}{\chiinf^{ig}}  
\, = \,
\frac{\langle N_{\mathcal{D}}^2 \rangle - \langle N_\mathcal{D} \rangle^2}{\langle N_\mathcal{D} \rangle} ,
\label{eq:bla}
\end{equation}
holds not only when $\mathcal{D}$ is large, but also, under the proviso that
this latter volume is negligible compared to the whole available space.
In that case, $\chi_{_L}(\mathcal{D}) = \chiinf$.
In practice of course, one has to work with a finite system, taken to be a cubic box of
length $L$, so that the condition for the validity of \eqref{eq:bla} reads
$\sigma^d \ll  \mathcal{A}(\mathcal{D}) \ll L^d$, with $\sigma$ some microscopic length.
This emulates grand canonical 
environment. It is thus important to know how $\chi_{_L}(\mathcal{D})$ defined in \eqref{eq:chiL}
depends both on $\mathcal{D}$ and $L$. We will see that this dependence is universal. Before going into the details of the measure of finite size effects, we present here the model used throughout this paper as a  test-bench and prototypical interacting fluid. 

\begin{figure}[h]
\begin{center}
\includegraphics[width=0.4\columnwidth,height=0.29\columnwidth,clip=true]{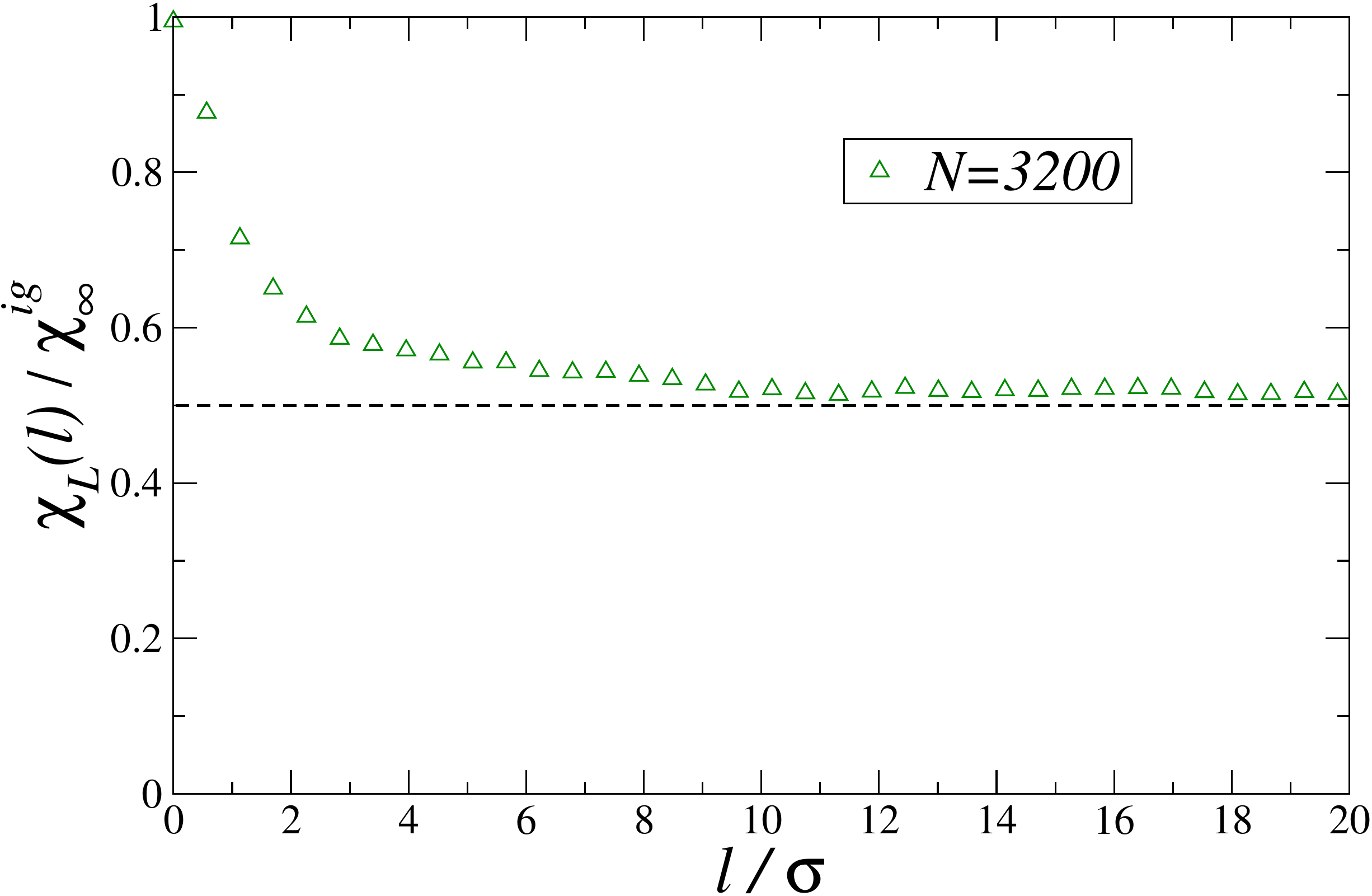}
\includegraphics[width=0.4\columnwidth,clip=true]{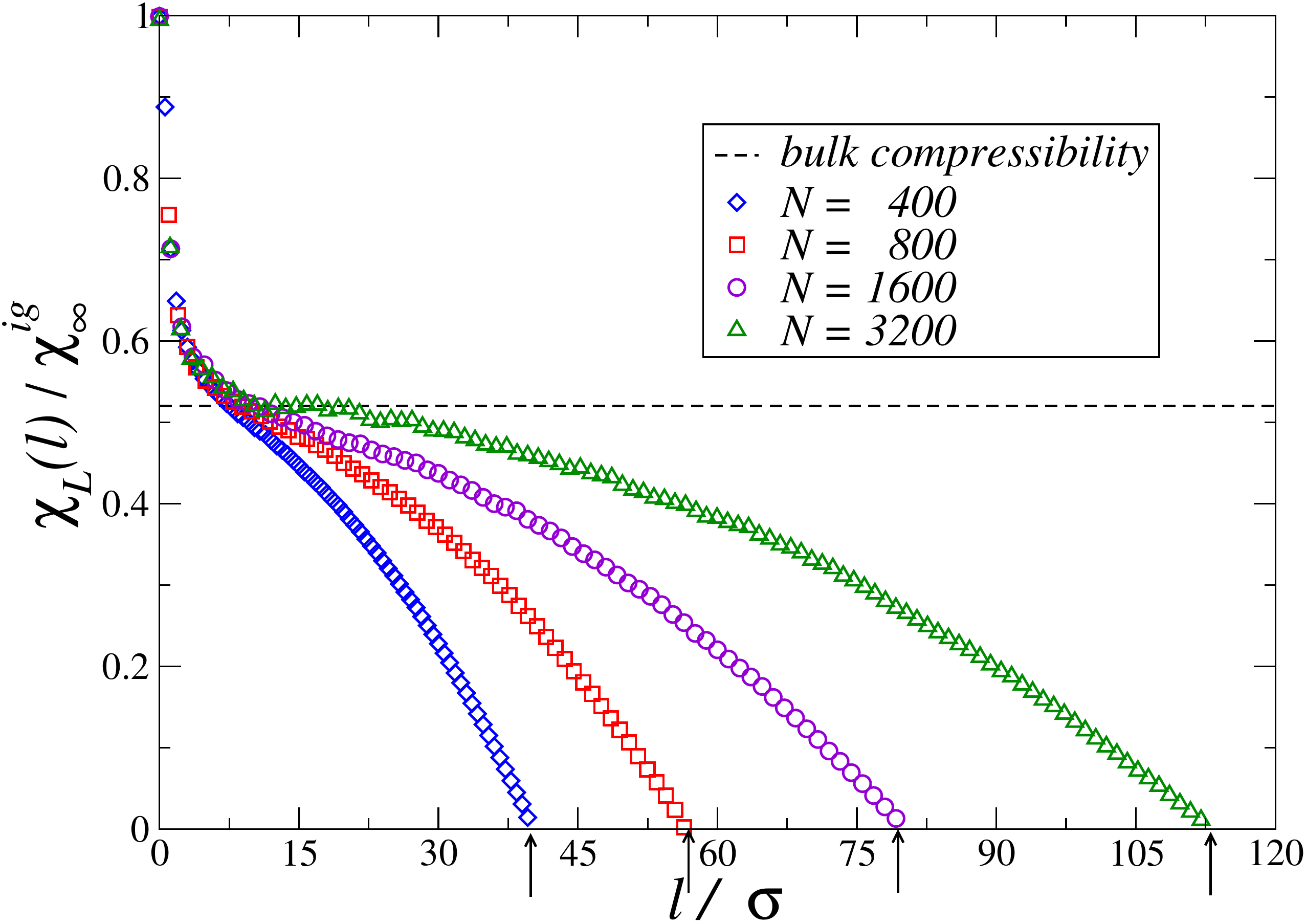}

\caption{\label{fig:compr_inter} Molecular dynamics results for a cubic control volume $\mathcal{D}$ of linear extension $l$, in a simulation box of size $L$. Left: behavior for small $l$ of the
  reduced variance defined in (\ref{eq:chiL}). Right: same quantity
  for different numbers of particles.
  In both figures, the reduced density ($\rho^*\equiv \rho \sigma^2$ where $\sigma$ is particle size) 
  is fixed to  $\rho^{*}=0.26$, and the horizontal dashed line shows
  $\chiinf/\chiinf^{ig}$. For each value of $N$, the vertical arrow shows $L/\sigma \equiv (N/\rho^*)^{1/2}$.}
\end{center}
\begin{pspicture}(0,0)(0.5,0.1)
\end{pspicture}
\vspace{-0.8cm}
\end{figure}

We have considered
$N$ particles moving in a $2$-dimensional box of length $L$ with periodic boundary conditions. 
The highly repulsive interaction potential $U$ is pair-wise additive and taken of the form:
\begin{equation}
U({\bm r}_{i},{\bm r}_{j})\propto \left(\frac{\sigma}{\vert {\mathbf r}_{i}-{\mathbf
    r}_{j}\vert}\right)^{64}.
\end{equation}
With such an interaction, the phenomenology
is very close to the hard disks case, with an effective diameter
$\sigma_{\text{eff}}\simeq \sigma$ (for instance, the bulk compressibility is well estimated by the Henderson's equation for hard disks~\cite{H75}). Therefore, we introduce the reduced effective
density, defined as
\begin{equation}
\rho^{*} \,=\, \frac{N}{L^{2}} \, \sigma^{2}, 
\label{density}
\end{equation}
which measures the relevance of interaction among the particles.  The
ideal gas is retrieved in the limit $\rho^{*} \to 0$. The system has
been studied via molecular dynamics simulations at fixed energy,
namely in the ensemble NVE.~\cite{AT89,HM06,FrenkelSmit} In short, the
equations of motion are numerically integrated via the velocity Verlet
algorithm with periodic boundary conditions, assuring
homogeneity.~\cite{AT89,HM06,FrenkelSmit}. In order to study
finite size effects, we have then measured the number of particles
variance and mean in a square domain $\mathcal{D}$ of size $l$.

\begin{figure}[hbt] 
\begin{center}
\includegraphics[width=.5\columnwidth,clip=true]{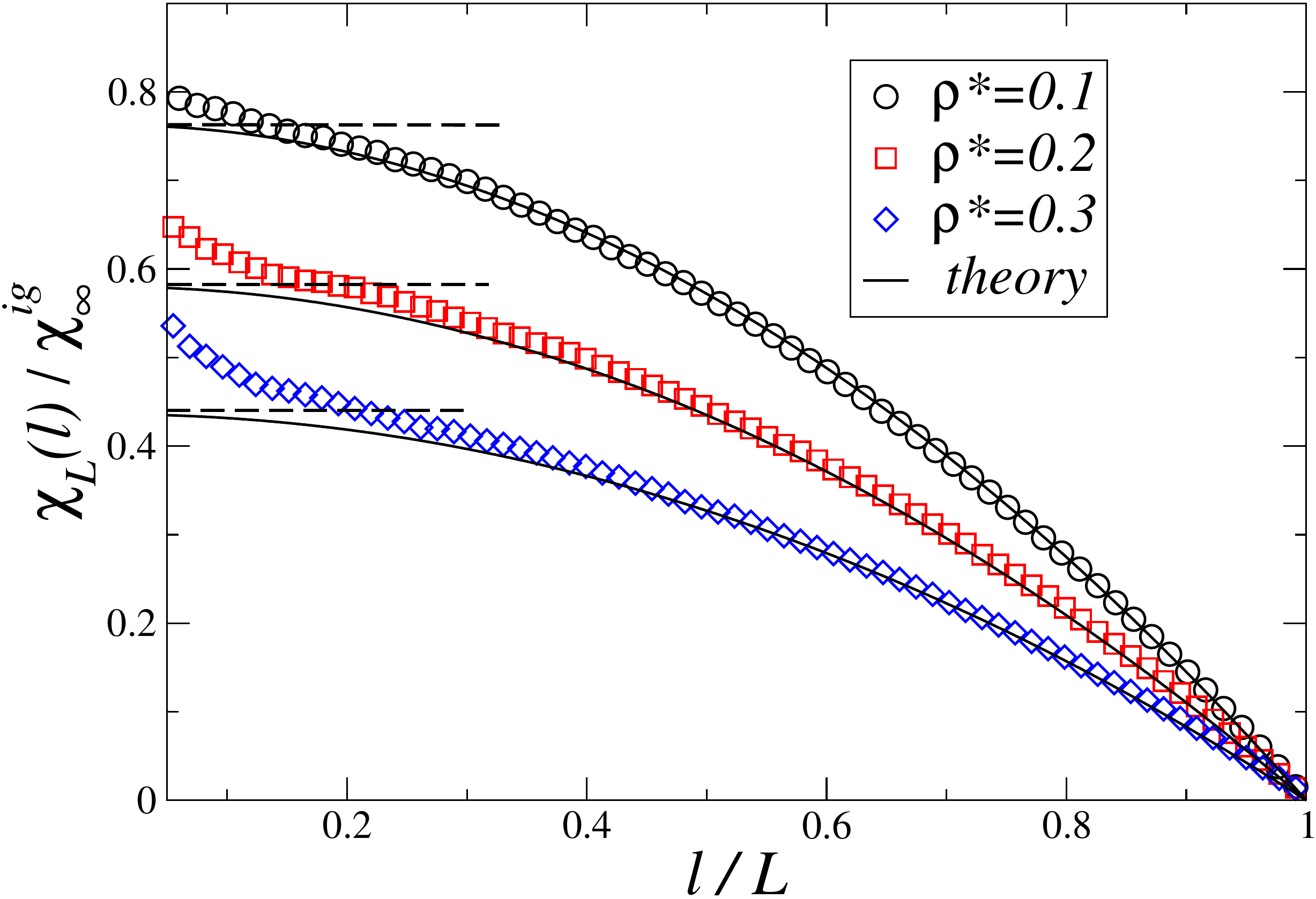}
\caption{\label{fig:inside_box} Behavior of the reduced variance 
defined in \eqref{eq:chiL}, which can arguably be viewed as a finite size compressibility,
  for the case $l<L$ and different effective densities. As above, $L$ is the simulation box size
  and $l$ the size of the control volume $\cal D$, inside which the number of particle $N_{\cal D}$ fluctuates.
  When $l=L$, $N_{\cal D}=N$, the known number of particles in the simulation box, and thus does not
  fluctuate ($\chi_{_L}(L)=0$).
  The solid line is
  the prediction given by Eq.~\eqref{eq:factoriz}. The horizontal dashed curves on the left
  hand side show the ratio $\chiinf/\chiinf^{ig}$ of the bulk compressibility
  over its ideal gas counterpart.}
\end{center}
\end{figure}

On Fig \ref{fig:compr_inter}-left, we see that $\chi_{_L}(\mathcal{D})$
indeed goes to a plateau for $l\gg\sigma$, giving the thermodynamic
compressibility of the system. This is true however provided $l\ll L$,
and when $l/L$ is no longer small, $\chi_{_L}(l)$ decreases
strongly, see Fig \ref{fig:compr_inter}-right. In the limiting case where
$l=L$, the number of particles in the control volume no longer fluctuates since it takes the known value $N$. Hence  $\chi_{_L}(L)=0$
as can be seen on the figure. More specifically, it has been shown that~\cite{SPANISH_AJP}
\begin{equation}
\frac{\chi_{_L}(l)}{\chiinf^{ig}} \,=\, \frac{\chiinf}{\chiinf^{ig}}\left(1-\frac{l^{2}}{L^{2}}\right),
\label{eq:factoriz}
\end{equation}
under the proviso that $l \gg \sigma$ i.e. that one should not probe
microscopic control volumes. 
In the present geometry, this relation is well obeyed, see Fig~\ref{fig:inside_box}.
More generally, when $\mathcal{D}$ is arbitrary and of volume
$\mathcal{A}(\mathcal{D})$,
the parenthesis reads $\left(1-\mathcal{A}(\mathcal{D})/L^{d}\right)$ in arbitrary dimension $d$. It therefore appears that the finite-size effects, embedded in the term in parenthesis, factorize from those due to interactions (given by $\chiinf/\chiinf^{ig}$).
We will soon offer below a derivation in two steps of Eq.~\eqref{eq:factoriz}
that
differs from existing ones, first when $l$ is close to $L$, where the factorization property is immediately apparent, and then without restrictions but for $l\gg \sigma$. The situation  where $l>L$ will be the subject of Sec.~\ref{sec:outside}. It is noteworthy to stress that in the ideal gas case, 
the relation $\chi_{_L}^{ig}(l)/\chiinf^{ig}=(1-l^2/L^{2})$ follows from elementary considerations, see Appendix~\ref{app:B}
which repeats the main arguments used in Ref.~\cite{SPANISH_AJP}.

At this point, one may wonder about the possible finite size effects that may affect the 
right hand side of Eq.~\eqref{eq:comp}. This equality does not assume $l\ll L$, but it requires 
a large enough domain $\cal D$ ($l \gg \sigma$). Interestingly, it may be extended to hold for
arbitrary $\cal D$: in general, we have 
\begin{equation}
\frac{\langle N_{\mathcal{D}}^2 \rangle - \langle N_\mathcal{D} \rangle^2}{\langle N_\mathcal{D} \rangle} 
\,=\, 
1 + \rho \int_\mathcal{\widetilde{D}} w_{\mathcal{D}}(\bm{r})\left[ g_{N}(\bm{r})-1 
\right] \, d \bm{r} ,
\label{eq:status}
\end{equation}
the derivation of which can be found in Appendix \ref{app:compr}.
As compared to \eqref{eq:comp}, note the weighting factor $w_{\mathcal{D}}$ and 
integration over a different domain $\widetilde{\mathcal{D}}$. 
When the volume $\mathcal{D}$ is large enough, the right hand side equality of \eqref{eq:comp}
holds, as explained in Appendix \ref{app:compr}. 

\begin{figure}[hbt] 
\begin{center}
\includegraphics[width=.5\columnwidth,clip=true]{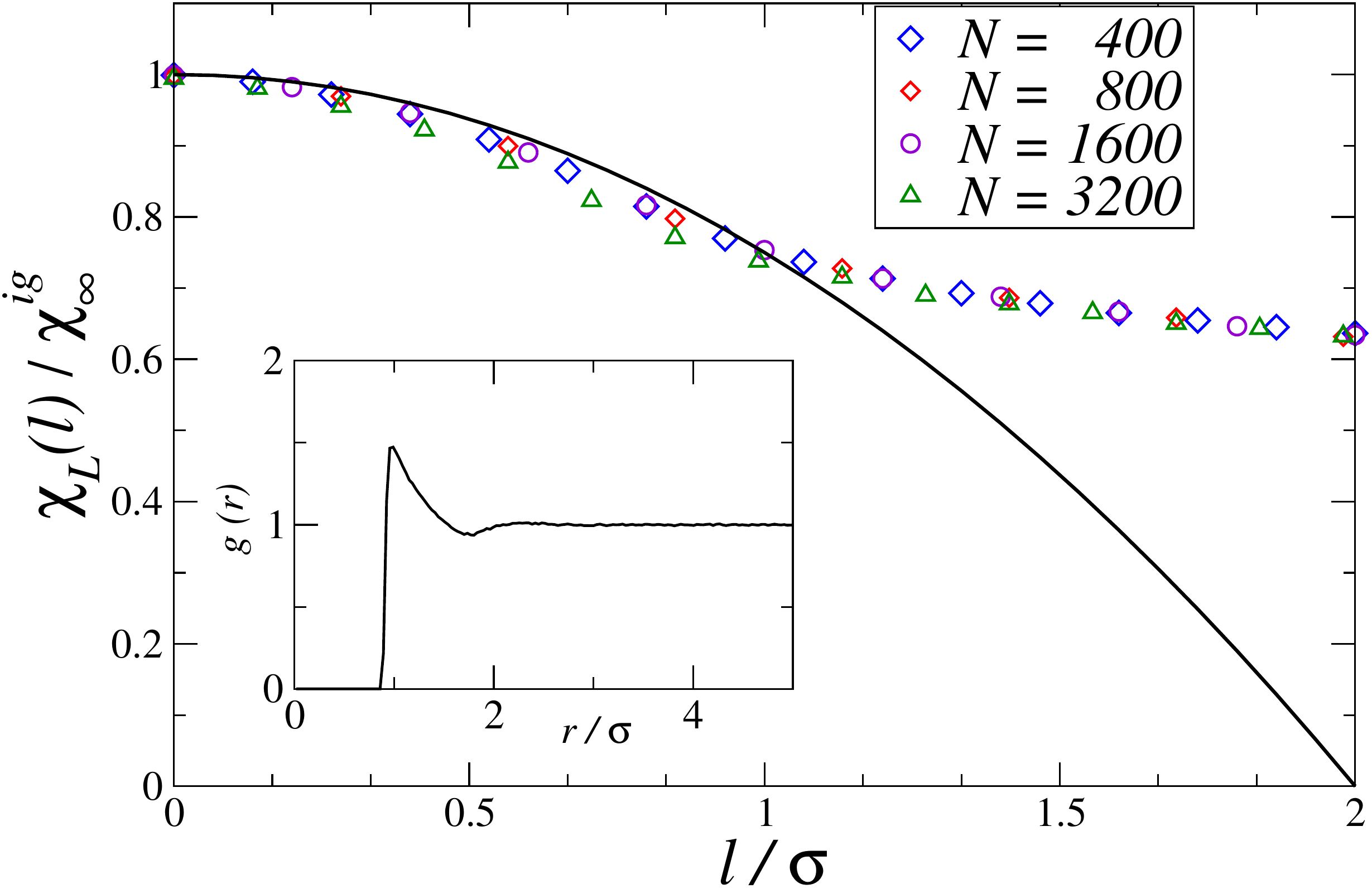}
\caption{\label{fig:microzoom} Same as Fig. \ref{fig:compr_inter}, zooming into
the microscopic region. The simulation data collapse onto the thick line,
showing the prediction of Eq. (\ref{eq:microchi}). This regime cannot be seen
in Fig. \ref{fig:inside_box} since the $x$ axis does not start at 0 there.
Inset: plot of the corresponding pair correlation function $g(r)$, where it
appears that $g\simeq 0$ for $r<\sigma$, as a consequence of strong 
interparticle repulsion.}
\end{center}
\end{figure}

By means of Eq.~\eqref{eq:status} it is possible to explain the behavior of the finite size compressibility $\chi_{_L}(\mathcal{D})$ for short length scales. Indeed, we come back to Fig.~\ref{fig:compr_inter}
and focus on the microscopic range where $l$ is comparable to the
particle size $\sigma$, and where $\chiinf/\chiinf^{ig}$ raises to
unity. First, this can be readily understood from the limiting case where $l$
is quite smaller than $\sigma$, so that there is at most one particle
in $\mathcal{D}$, with a small probability $p\ll 1$, or none with
probability $1-p$. Hence $N_{\mathcal{D}}$ becomes here a Bernoulli
variable ($N_{\mathcal{D}}=0$ or $1$), so that the variance
$\left<N^{2}_{\mathcal{D}}\right>-\left<N_{\mathcal{D}}\right>^{2}=p-p^{2}\simeq
p$, which thus equals $\left<N_{\mathcal{D}}\right>$. The left-hand
term of Eq.~\eqref{eq:status} is accordingly unity and so is the right-hand side,
where the integral is small compared to $1$. When the
volume of $\mathcal{D}$ increases, this integral contributes
negatively since $g(r)\simeq 0$ for $r<\sigma$, as a consequence of
hard core repulsion. This explain the initial decay observed for small
sizes in Figs.~\ref{fig:compr_inter}, which can be further rationalized 
since having a vanishing $g$ for small inter-particle distances implies, from 
(\ref{eq:status}), that
\begin{equation}
\frac{\chi_{_L}(l)}{\chiinf^{ig}} \,\simeq \, 1- \rho \,l^2 =
1- \rho^* \left(
\frac{l}{\sigma}\right)^2 \quad \hbox{for} \quad l < \sigma ,
\label{eq:microchi}
\end{equation}
where use has been made of of $\int_{\widetilde{\mathcal{D}}} w_{\mathcal{D}}(\bm{r}) d {\bm{r}} = \mathcal{A}(\mathcal{D})$.
As can be seen in Fig.~\ref{fig:microzoom}, which also displays a typical $g({\bm r})$, this relation is well obeyed.

\subsection{Factorization properties and fluctuations in a large sub-region} 
\label{sec:granc}

We know that $\chi_{_L}(L)=0$, and we give here a simple argument which
explains how $\chi_{_L}(l)$ approaches $0$ when $l$ approaches $L$,
following the form given in Eq. \eqref{eq:factoriz}. Space dimension $d$
is here unspecified.
We consider again a square sub-domain $\mathcal{D}$ (size $l$) of the confining
box (size $L$). When $l$
is close to $L$ (with $l<L$), the idea is to consider the
complementary domain of $\mathcal{D}$, noted $\overline{\mathcal{D}}$,
such that the volumes obey
$\mathcal{A}(\mathcal{D})+\mathcal{A}(\overline{\mathcal{D})}=L^{d}$. We
assume that $\overline{\mathcal{D}}$, although much smaller than
$\mathcal{D}$, is large enough compared to the microscopic scale $\sigma$. Under these circumstances, $\mathcal{D}$ plays for $\overline{\mathcal{D}}$ the role of a reservoir of particles, and $\overline{\mathcal{D}}$ is ruled by the grand canonical laws. In particular, the number of particles in $\overline{\mathcal{D}}$, $N_{\overline{\mathcal{D}}}$, obeys
\begin{equation}
\frac{\chiinf}{\chiinf^{ig}}  
\, = \,
\frac{\langle N_{\overline{\mathcal{D}}}^2 \rangle - \langle N_{\overline{\mathcal{D}}} \rangle^2}{\langle N_{\overline{\mathcal{D}}} \rangle}.
\end{equation}
We then note that $N_{\mathcal{D}}+N_{\overline{\mathcal{D}}}\,=\,N$ is a
non fluctuating quantity. Hence, $N_{\mathcal{D}}$ and
$N_{\overline{\mathcal{D}}}$ have the same variance, a quantity that
will be denoted by $V$:
\begin{equation}
V(N_{\DD})=V(N_{\OD})
\quad \Longrightarrow \quad
\left<N^{2}_{\DD}\right>-\left<N_{\DD}\right>^{2}\,=\, \left<N^{2}_{\OD}\right>-\left<N_{\OD}\right>^{2}.
\end{equation}
On the other hand, we have from homogeneity
\begin{equation}
\left<N_{\DD}\right>=\frac{\mathcal{A}(\mathcal{D})}{L^{d}} N \quad \textrm{and} \quad \left<N_{\OD}\right>=\left(1-\frac{\mathcal{A}(\mathcal{D})}{L^{d}}\right) N.
\end{equation}
Gathering results, we reach the desired expression for $\chi_{_L}(l)$
\begin{equation}
  \frac{\chi_{_L}(l)}{\chiinf^{ig}}  
  \, = \,
  \frac{\langle N_{\mathcal{D}}^2 \rangle - \langle N_\mathcal{D} \rangle^2}{\langle N_\mathcal{D} \rangle}
  \,=\,
  \frac{\chiinf}{\chiinf^{ig}}\frac{L^{d}-\mathcal{A}({\mathcal{D}})}{\mathcal{A}(\mathcal{D})}\simeq 
  \frac{\chiinf}{\chiinf^{ig}}\left( 1-\frac{\mathcal{A}({\mathcal{D}})}{L^{d}}\right)
  =  \frac{\chiinf}{\chiinf^{ig}}\left( 1-\frac{l^d}{L^{d}}\right),
  \label{fact}
\end{equation}
where use was made of the condition
$\mathcal{A}(\mathcal{D})\simeq L^{d}$. As expected,	
$\chi_{_L}(L)=0$. We have thus justified that finite size effects arise
in $\chi_{_L}(l)$ through a purely geometric factor, 
$( 1-\mathcal{A}({\mathcal{D}})/L^{d})$, a result that so far holds
under the requirement that $\DD$ almost fills the available volume $L^d$. 
As shown in Fig.~\ref{fig:inside_box}, it is possible to extend the validity of that expression in a large range of $\DD$
volumes, provided $\mathcal{A}({\mathcal{D}})\gg \sigma^{d}$, in order to wash
out microscopic details. This is the purpose of the next section.

\subsection{Factorization property and pair correlation function}
\label{ssec:factog}

The line of reasoning goes here through the particle correlation
function $g_{N}({\bm r})$ defined in Appendix~\ref{app:A}. We remind
that $\rho g_{N}({\bm r})$ is the density of molecules at point ${\bm r}$, given that one molecule is at the origin.
Its expression is slightly different in a finite box with $N$
molecules, and in a truly infinite system at the same density, where
it is denoted by $g({\bm r})$. In order to get a flavor of the
leading order difference between $g_{N}({\bm r})$ and $g({\bm r})$, we
go back to the ideal gas case, for which all the $g$s have to be
uniform, since there is no length scale in the model. By definition,
$\rho \int g_{N}({\bm r})$ counts the number of neighbors around a
given tagged molecule, which is therefore $N-1$. This means that
$g_{N}({\bm r})=1-1/N$, which in turn implies that $g({\bm r})=1$.~\cite{HM06} 

Instructed by the ideal gas limiting case where $g_{N}({\bm r})=g({\bm r})-1/N$,
we go back to interacting systems and assume that the first finite $N$ correction to $g({\bm r})$ reads
\begin{equation}
g_{N}({\bm r})= g({\bm r}) + \delta g \label{eq:gr_exp}
\end{equation}
where $\delta g$ does not depend on ${\bm r}$, and
is likely to scale like $1/N$.
 We then use Eq.~\eqref{eq:status} for a non microscopic domain $\mathcal{D}$, in which case
\begin{equation}
\frac{\chi_{_L}(l)}{\chiinf^{ig}}  
\, \equiv \,
\frac{\langle N_{\mathcal{D}}^2 \rangle - \langle N_\mathcal{D} \rangle^2}{\langle N_\mathcal{D} \rangle} 
\,=\, 
1 + \rho \int_\mathcal{D} \left[ g_{N}(\bm{r})-1 
\right] \, d \bm{r} .
\end{equation}
From Eq.~\eqref{eq:gr_exp} we have
\begin{eqnarray}
  \frac{\chi_{_L}(l)}{\chiinf^{ig}}&=&1 \,+\,\rho\int_{\mathcal{D}} (g({\bm r}) -1 ) d {\bm r} \,+\, \rho \int_{\mathcal{D}}\delta g d{\bm r}\nonumber\\
&=&\frac{\chi_{\infty}}{\chiinf^{ig}}+\rho \mathcal{A}(\mathcal{D}) \,\delta g, \label{chi_from_gdr}
\end{eqnarray}
where Eq.~\eqref{eq:comp} has been used.\\
Next, a necessary requirement is that $\chi_{_L}(L)=0$, which imposes that
\begin{equation}
\rho \,\delta g \,L^{d}\,=\, -\frac{\chiinf}{\chiinf^{ig}}
\quad \Longrightarrow \quad
\delta g = - \frac{\chiinf}{\chiinf^{ig}} \, \frac{1}{N}. \label{expr:dg}
\end{equation}
We obtain here a non-trivial result, which generalizes to interacting systems the ideal gas relation (see also~\cite{LP61bis} for a more general discussion):
\begin{equation}
g_{N}({\bm r})=g({\bm r})-\frac{\chiinf}{N \chiinf^{ig}}.\label{correction}
\end{equation}
 Inserting Eq.~(\ref{expr:dg}) into
Eq.~(\ref{chi_from_gdr}), one obtains
\begin{equation}
  \frac{\chi_{_L}({\cal D})}{\chiinf^{ig}} \,=\,
  \frac{\chiinf}{\chiinf^{ig}}
  \left( 1-\frac{\mathcal{A}({\mathcal{D}})}{L^{d}}\right), \label{factbis}
\end{equation}
which is exactly Eq.~(\ref{eq:factoriz}). We point out that there is also another term of order $\mathcal{O}(1/N)$ in Eq.~\eqref{correction}. However, it 
depends on ${\bm r}$ and contributes to the compressibility only with a sub-leading term in system size, leaving the result 
in Eq.~\eqref{factbis} unaffected since the domain $\mathcal{D}$ is here taken large
(see~\cite{RWV97} for a detailed discussion).
We thus see that the ideal gas finite-size correction  (\ref{ideal_gas_compress}) 
also applies to interacting fluids.
It is also noteworthy to stress that we did not assume, at any stage, 
that the interaction potential was pair-wise additive (except in the
simulated model used for illustrative purposes).

The goal in the subsequent analysis is to decipher the fluctuation behavior of the number of molecules
$N_{\cal D}$, through $\chi_{_L}(l)$, when the control volume is larger than the confining box
($l>L$, with periodic boundary conditions). Clearly,  Eq.~(\ref{factbis}) no longer holds ``outside the box'', where ${\mathcal{A}(\mathcal{D})>L^{d}}$, since it would predict 
a negative variance for $N_{\cal D}$.

\section{Fluctuations outside the box}
\label{sec:outside}

\begin{figure}[h!bt]
\begin{center}
 \includegraphics[width=.7\columnwidth,clip=true]{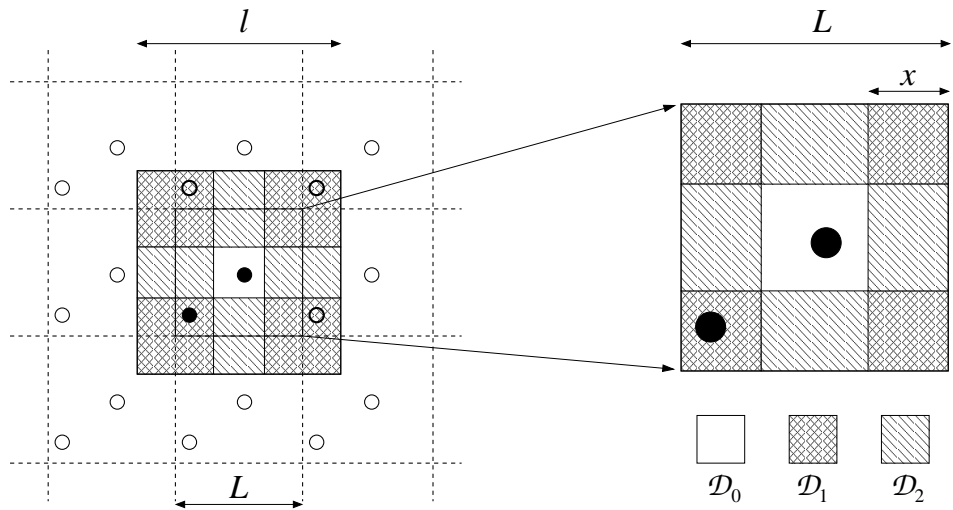}
\caption{\label{fig1} Left: graphical representation of the confining (simulation) box of size $L$ 
  with periodic
  boundary conditions in the two dimensional case, with $L<l<2L$. This means that $\kappa=1$ here,
  see the definition~\eqref{kappa_def}. Since a control volume
  $\mathcal{D}$  larger than the confining box is
  considered,  the relevant images of the particles must  be
  taken into account.  Right: zoom defining the regions
  $\mathcal{D}_{0}$, $\mathcal{D}_{1}$ and $\mathcal{D}_{2}$ inside the simulation box,
  for the computation of the
  fluctuations in Eq.~\eqref{fluct_eq}. Here, the simulation box contains two particles 
  (the filled circles), one belonging to ${\cal D}_0$ (so that $N_0=1$) and one to ${\cal D}_1$
  (and thus $N_1=1$). There are consequently 5 particles in the control volume (square of size $l$),
  in agreement with Eq. (\ref{decomposition}), since $N_2=0$ here.
}
\end{center}
\end{figure}

If one consider domains larger than the
simulation box, due account must be taken of the presence of the images in the replicated boxes 
of a given particle
from the simulation box,  in order to have a
proper calculation of the fluctuations. For the sake of simplicity, we
discuss again the two dimensional case with a square region
$\mathcal{D}$. We postpone the discussion on the
relevance of the dimensionality to the end of the section. Let us
consider the representation in Fig.~\ref{fig1} (left), where $\cal D$ is of length $l$ such that $L<l<2L$. The number $N_{t}$ of
particles contained in this control volume can be so
decomposed:
\begin{equation}
N_{t}=N_{0}+4 N_{1}+2 N_{2}, \label{decomposition}
\end{equation}
where $N_{1}$ and $N_{2}$ are respectively the number of particles
contained in the regions $\mathcal{D}_{1}$ and $\mathcal{D}_{2}$
showed in Fig.~\ref{fig1} (right) and $N_{0}$ is the number in the
central white area $\mathcal{D}_{0}$. Since the number of particles
in the system is fixed, these quantities obey the following
constraint:
\begin{equation}
N=N_{0}+ N_{1}+N_{2}, \label{constraint}
\end{equation}
which is therefore non-fluctuating. Eq.~\eqref{constraint} can be used to simplify Eq.~\eqref{decomposition}
since the the confining box is entirely contained into the region of length $l$ and does not contribute
to the fluctuations. Then, the total number of particle inside $\mathcal{D}$,
from Eq.~(\ref{decomposition}), can be rewritten as $N_{t}=N+ N_{fl}$ where 
\begin{equation}
N_{fl}\equiv 3N_{1}+N_{2}, \label{N_fluct}
\end{equation}
and both quantities have the same variance:
$V(N_{fl}) = V(N_t)$.

\begin{figure}[h!bt] 
\begin{center}
\includegraphics[width=.5\columnwidth,clip=true]{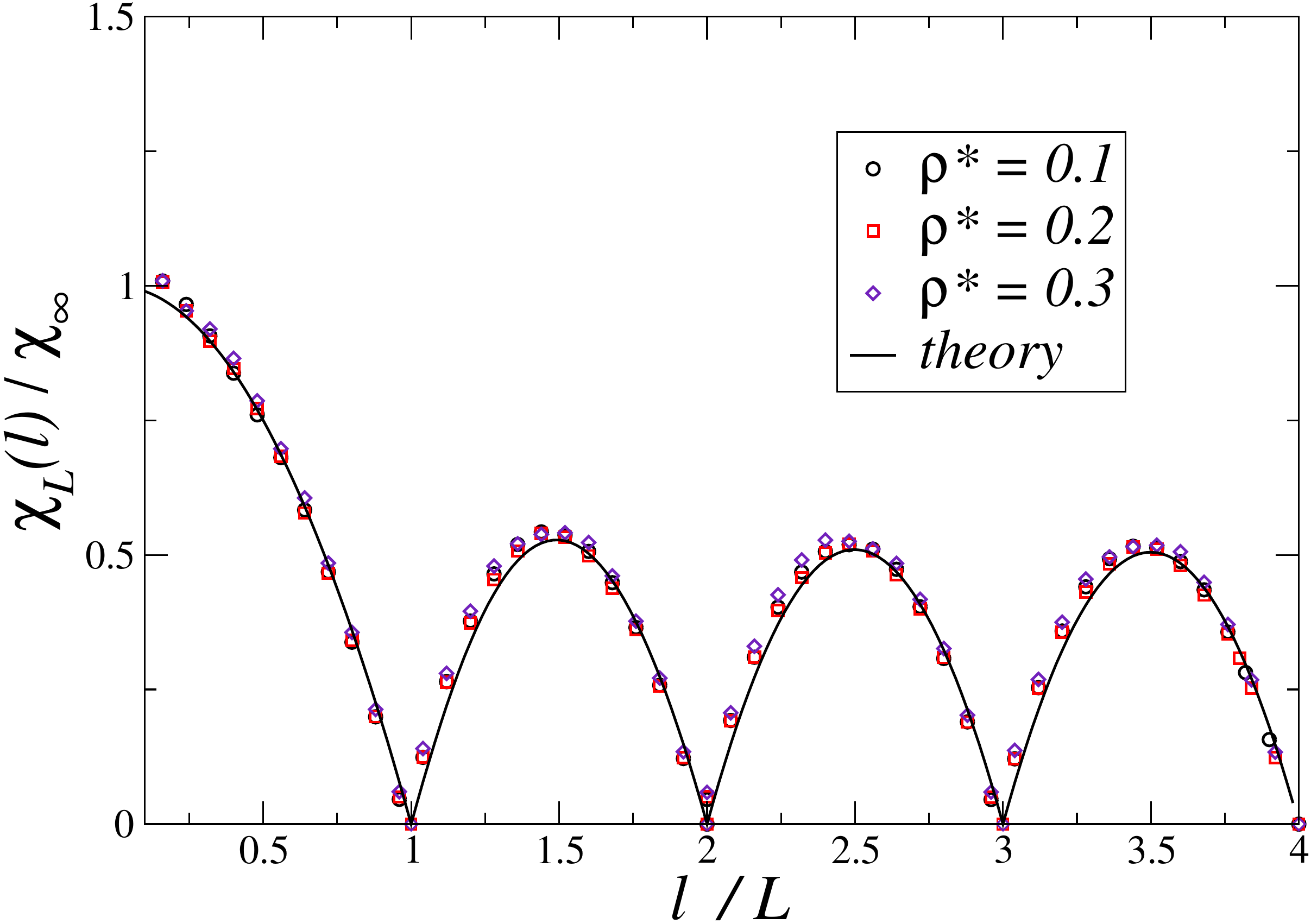}
\caption{\label{fig:interaction_compress}Rescaled compressibility ``outside'' the confining box for the quasi hard disk system at various densities, and comparison with Eq.~\eqref{compr_2d}, shown by the continuous line.  The bulk compressibility $\chiinf$ depends on $\rho^{*}$.}
\end{center}
\end{figure}

From this example one learns that, in order to describe the number
fluctuations for every length $l$, an important quantity is the number
of cells (the original one and the replicated ones) which are included
in the domain of interest. Therefore, the relevant parameters are
\begin{eqnarray}
\kappa=\left[\frac{l}{L}\right]\qquad \textrm{and } \qquad x=\frac{l - \kappa L}{2}, \label{kappa_def}
\end{eqnarray}
where $[\dots]$ is the ``integer part'' function~\footnote{$[x]=n$
  where $n$ is the largest integer that satisfies the inequality $n\le
  x$.}. Note that $x< L/2$, by definition.
The relation (\ref{N_fluct}) can
be easily generalized to any value of $l$, also larger than $2L$, yielding:
\begin{equation}
N_{fl}=(2\kappa +1) N_{1}+\kappa N_{2}. \label{fluct_eq}
\end{equation}
In conclusion, it appears from Eq.~(\ref{fluct_eq}) that, due
to periodicity, the particle fluctuations ``outside'' the box can be
recast as a sum of contributions ``inside'' the confining box. We can therefore resort to the considerations of Appendix \ref{sec:app}, 
that rely on the so-called law of total variance. It is shown there that
\begin{equation}
V(\alpha N_{1}+\beta N_{2}) \, =\, 
\frac{(\alpha p_{1}+\beta p_{2})^{2}}{p_{1}+p_{2}} V(N_{1}+N_{2}) \, + \, (\alpha-\beta)^{2} \,E\left[ V(N_{1}\vert N_{1}+N_{2}) \right],
\label{eq:interm11}
\end{equation}
where in the last term $V(N_{1}\vert N_{1}+N_{2}) =V(N_{2}\vert N_{1}+N_{2}) $ 
is the variance of $N_{1}$, given that $N_{1}+N_{2}$ has a
prescribed fixed value. Once this variance is known for fixed $N_{1}+N_{2}$, it should subsequently 
be averaged over the probability distribution of $N_1+N_2$, to provide the mean
$E\left[ V(N_{1}\vert (N_{1}+N_{2})) \right]$ sought for, appearing on the right hand-side of
(\ref{eq:interm11}).
The key point next is that we can apply the compressibility relation (\ref{eq:factoriz}) twice, to get first $V\left(N_{1}+N_{2}\right)$ and then $V\left(N_{1}\vert (N_{1}+N_{2})\right)$. In the present geometry,
\begin{equation}
p_{1}=\frac{\mathcal{A}(\mathcal{D}_{1})}{L^{2}}=\frac{4x^{2}}{L^{2}} \ ,\qquad \qquad p_{2}=\frac{\mathcal{A}(\mathcal{D}_{2})}{L^{2}}=\frac{4x(L-2x)}{L^{2}},\label{eq:geometry}
\end{equation}
but for the time being, we do not need to specify these values.
To get $V(N_{1}+N_{2})$, we consider the region $\mathcal{D}_{1}\cup \mathcal{D}_{2}$, which is a sub domain of the confining box having $N$ particles, so that
\begin{equation}
V(N_{1}+N_{2})=\frac{\chiinf}{\chiinf^{ig}}\left<N_{1}+N_{2}\right>\left(1-\frac{\AD{1}+\AD{2}}{L^{2}}\right)
=\frac{\chiinf}{\chiinf^{ig}} N (p_{1}+p_{2})(1- p_{1}-p_{2}).
\end{equation}
Likewise $V(N_{1}\vert N_{1}+N_{2})$ is the variance of the number of particles in $\AD{1}$, given that there are exactly $N_{1}+N_2$ particles in $\mathcal{D}_{1}\cup \mathcal{D}_{2}$. Thus
\begin{equation}
V(N_{1}\vert N_{1}+N_{2})= \frac{\chiinf}{\chiinf^{ig}}\, E(N_{1}\vert (N_{1}+N_{2})) \left(1-\frac{\AD{1}}{\AD{1}+\AD{2}}\right)=
\frac{\chiinf}{\chiinf^{ig}} \left< N_1+N_2\right>\frac{p_1}{p_1+p_2}\left(1-\frac{\AD{1}}{\AD{1}+\AD{2}}\right)
\end{equation}
and we finally need $E(N_{1}+N_{2})=\left<N_{1}+N_{2}\right>= N(p_1+p_2)$ to reach
\begin{equation}
E\left[V(N_1\vert N_1+N_2)\right]= \frac{\chiinf}{\chiinf^{ig}} N p_1 \left(1-\frac{p_1}{p_1+p_2}\right).
\end{equation}
Going back to Eq. (\ref{eq:interm11}), this yields
\begin{equation}
V(N_{t})\, =\, \frac{\chiinf}{\chiinf^{ig}} \,
N (p_{1}+p_{2}) \left[\left(\frac{\alpha p_{1}+\beta p_{2}}{p_1+p_2}\right)^{2}(1-p_{1}- p_{2})+(\alpha-\beta)^{2}\left(\frac{p_{1}p_{2}}{\left(p_1+p_2\right)^{2}} \right)\right].
\end{equation}
This expression is  general. We now specify it for the probabilities given in Eq. (\ref{eq:geometry}), obtaining
\begin{equation}
V(N_{fl})\, =\, \frac{\chiinf}{\chiinf^{ig}} \,N\frac{4x (L-2x) \left(\kappa^2 L^2+4 \kappa
  Lx+x (L+2x)\right)}{L^4}. \label{variance_2d}
\end{equation}
Moreover, from Eqs.~\eqref{kappa_def}, the mean number of
particles in a box of length $l$ can be rewritten as:
\begin{equation}
  E(N_{t}) = \left<N_t \right> =\frac{N}{L^{2}}(\kappa L+2x)^{2}.\label{N_av}
\end{equation}
The final formula of the compressibility is then given by
\begin{equation}
   \frac{\chi_{_L}(l)}{\chiinf^{ig}} \,=\,
  \frac{\chiinf}{\chiinf^{ig}} \,\frac{4 x (L-2x) \left(\kappa^2 L^2+4 \kappa
  Lx+x (L+2x)\right)}{L^2 (\kappa L+2x)^{2}}, \label{compr_2d}
\end{equation}
which is remarkably well obeyed by simulation data, see
Fig.~\ref{fig:interaction_compress}. Let us note that in the case
$l<L$, where $\kappa=0$ and $x\equiv l$, Eq.~\eqref{factbis} is
recovered.  In the literature a distinction is 
  made between explicit (due to ensemble averages) and
  implicit (due to periodic boundary conditions) finite size effects.~\cite{pratt1} 
  It is then noteworthy that both effects are at work in Eq. \eqref{compr_2d}.
  We stress again that we are describing here the
  most general and model independent contribution and we are
  neglecting the effects due to the microscopic region, namely when
  $x$ (or $L-x$) is of the order of the correlation
  length (around $\sigma$ in our case). At this scale, periodic
  boundary conditions can give rise to other implicit corrections, that can be relevant in denser regimes.~\cite{pratt2,RWGV99}

\begin{figure}[h!bt]
\begin{center}
\includegraphics[width=.5\columnwidth,clip=true]{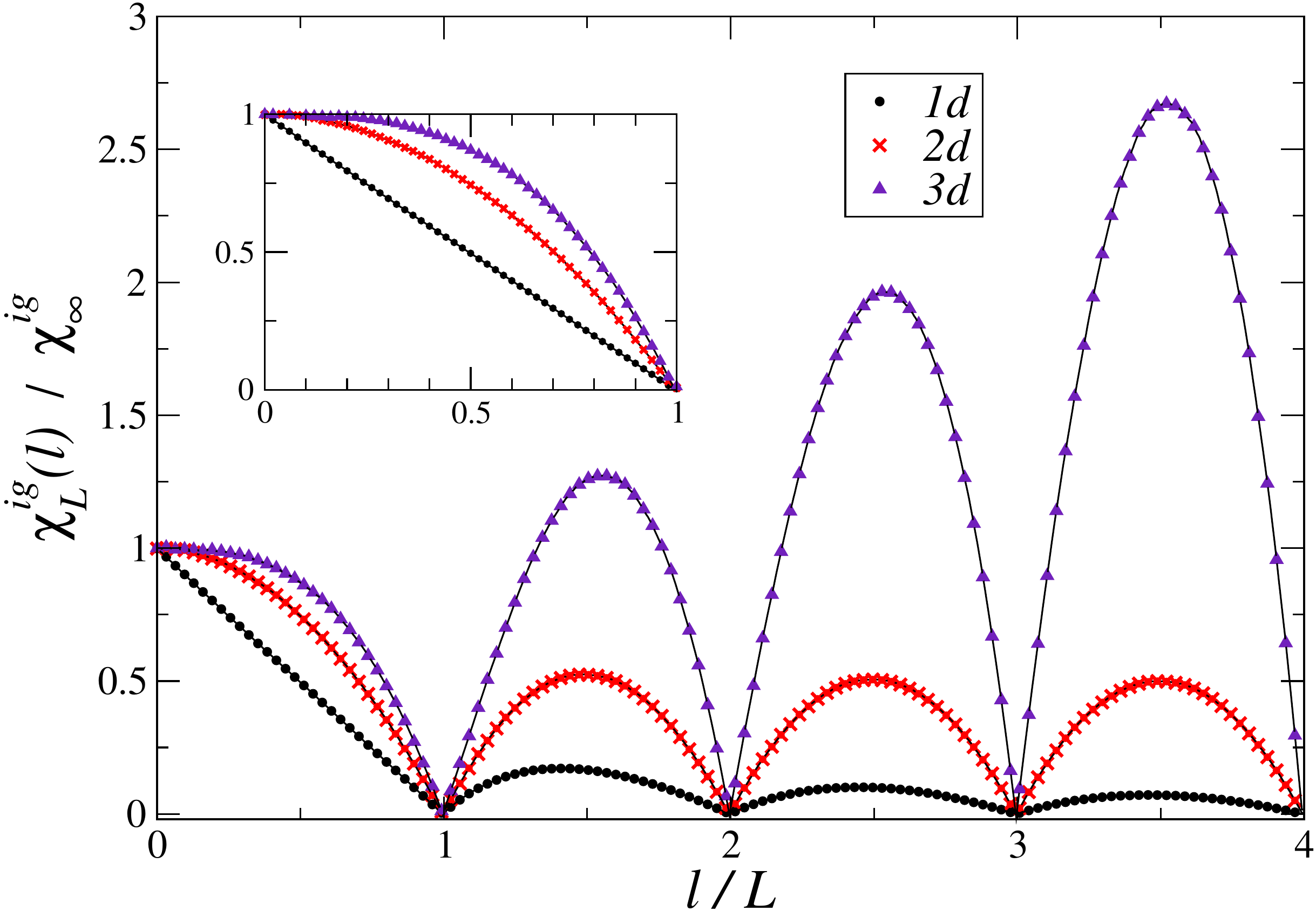}
\caption{\label{fig:dimensions} Numerical evaluation of $\chi_{_L}(l)$ 
in a non-interacting gas for $N=30$ particles, as a function of control volume
size, for dimensions $d=1,2,3$. The curves follow the
  predictions (see Eq.~\eqref{compr_2d} for the 2d
  case). Inset: Behavior of the compressibility for the first cell,
  where Eq~ (\ref{eq:factoriz}) (becoming (\ref{ideal_gas_compress})
  for the ideal gas) is obeyed.}
\end{center}
\end{figure}

The behavior of $\chi_{_L}(l)$ for $l>L$ is oscillatory, since $\chi_{_L}(l^{*})=0$
  for each $l^{*}$ that is commensurate with the length of the box,
  namely for $l^{*}=L,2L,3L\dots$. Then, for each interval $ \kappa L
  \le l \le (\kappa+1) L$,  $\chi_{_L}$ reaches a maximum at an $x$ value slightly
  lower than $x=L/4$, which tends asymptotically to $L/4$ for large
  $\kappa$. Indeed, for large $l$,
namely large $\kappa$, one has that:
\begin{equation}
\chi_{_L}^{(d=2)}(l)\simeq  \chiinf \frac{4x(L-2x)}{L^{2}}, \label{asimpt_compr2d}
\end{equation}
where the superscript keeps track of the dimensionality of the
system. Eq.~\eqref{asimpt_compr2d} has the interesting property of
being independent on $\kappa$, resulting asymptotically in a periodic
function. 

Finally, the
arguments presented in this section can be adapted
to different dimensions, and it turns out that the asymptotic behavior~\eqref{asimpt_compr2d} is
strongly connected to the dimensionality of the system. We shall not repeat the full
calculation, but we deduce instead the behavior for large $l$ from a simple
scaling approach. When $x\simeq L/4$, the fluctuation of
$N_{fl}$ stem from a surface contribution, and each different cell at the 
boundary of the domain under consideration behaves independently from the other ones.
The variances can therefore be added, resulting in
\begin{equation}
V(N_{fl})\propto \kappa^{2(d-1)} N \qquad\hbox{while}\qquad E(N_{t})\propto \kappa^{d} N. 
\label{eq:scaling}
\end{equation}
Eq.~(\ref{eq:scaling}) explains the peaks observed in Fig.~\ref{fig:dimensions}, since:
\begin{equation}
\chi_{_L}^{(d)}\propto \kappa^{d-2}  \qquad \textrm{for  $\kappa \gg 1$ and   $x\simeq \frac{L}{4}$},
\end{equation}
with therefore a growing amplitude for large $l$ in three dimensions, and conversely a
decrease in 1d.

\section{The structure factor}
\label{sec:structure}

\subsection{Definition and low $k$ behavior}

Having analyzed real space probes for density fluctuations and correlations
in sections \ref{sec:inside} and \ref{sec:outside}, we now turn to a related
and complementary
study in Fourier space. A convenient measure of density-density correlations,
used in experiments, theory and simulations alike, is given by the structure
factor, defined as
\begin{equation}
S({\bm k})=\frac{1}{N}\left<\widetilde{\rho}({\bm k})\widetilde{\rho}(-{\bm k})\right>, \label{Sk_factor}
\end{equation}
where the microscopic density $\widehat\rho$ and its Fourier transform are introduced:
\begin{equation}
\widehat{\rho}({\bm r})=\sum_{j=1}^{N} \delta({\bm r}-{\bm r}_{j}) 
\quad \Longrightarrow \quad
\widetilde{\rho}({\bm k})= \int d{\bm r} \widehat{\rho}({\bm r})e^{-i{\bm k} \cdot {\bm r}} =\sum_{j=1}^{N}e^{-i{\bm k} \cdot {\bm r}_{j}}.
\end{equation}
From Eq.~(\ref{Sk_factor}) follows that
\begin{equation}
S_L({\bm k}) \, =\, 
\frac{1}{N}\sum_{i=1}^{N}\sum_{j=1}^N\left<\exp\left[-i{\bm k} \cdot ({\bm r}_{i}-{\bm r}_{j})\right]\right>\nonumber\\
\, = \, 1+\frac{1}{N}\sum_{i\neq j}^{N}\left<\exp\left[-i{\bm k} \cdot ({\bm r}_{i}-{\bm r}_{j})\right]\right>,
\end{equation}
where the system size $L$ appears explicitly as a subscript.
For indiscernible particles, there are $N(N-1)$ possible couples $(i,j)$ involved in the summation above,
and thus 
\begin{equation}
 S_L({\bm k})
\, = \, 1+ (N-1) \left<\exp\left[-i{\bm k} \cdot {\bm r}_{12}\right]\right> .
\end{equation}
The equilibrium average $\left<...\right>$, which pertains to a homogeneous system, 
can be expressed in terms of $g_N(\bm r)$, but some
attention should be paid to normalization. Remembering that $\rho\int g_N(\bm r) d{\bm r} = N-1$
where the integral runs over the total available volume $L^d$ and $\rho = N/L^d$,
a simple requirement to set normalization right
is to enforce that $\left<1\right>$ be unity, so that 
\begin{equation}
S_L({\bm k}) \, = \, 1  +\frac{N}{L^{d}}\int d{\bm r} w_{L}(\bm{r})g_N({\bm r}) e^{-i{\bm k} \cdot {\bm r}}.
\label{ideal_gas_integral}
\end{equation}
where the weight function $w_{L}(\bm{r})$ is a consequence of changing
to the new variable ${\bm r}\equiv {\bm r}_{1}-{\bm r}_{2}$. It
already appeared in Eq.~\eqref{eq:status} and is defined in
Eq.~\eqref{gr_corr_square} of the Appendix~\ref{app:compr}. The term $w_{L}({\bm
  r})$ is quasi systematically overlooked in the literature, which assumes that $L$ is big enough, as we now assume. In that case, we do not need to specify the volume of integration in (\ref{ideal_gas_integral}), which avoids the subtleties discussed in Appendix~\ref{app:structure}. In the
thermodynamic limit ($L\to \infty$, $N\to \infty$, $N/L^d=\rho$
constant), one has
\begin{equation}
\lim_{L \to \infty} S_L({\bm k}) \equiv S_{\infty}({\bm k})=1+\rho \int_{\mathbb{R}^d}  d{\bm r} g({\bm r}) e^{-i{\bm k} \cdot {\bm r}}
\label{sk_asymptotic}
\end{equation}
which can be rewritten as:
\begin{equation}
S_\infty({\bm k}) \, = \, 
1 + (2\pi)^{d}\rho\delta({\bm k})+\rho \int_{\mathbb{R}^d} d{\bm r} \left[ g({\bm r})-1\right] e^{-i{\bm k} \cdot {\bm r}} .
\label{delta_in_k0}
\end{equation}
The quantity in the integral is quickly vanishing as $r=|\bm r|$ becomes
large, a property which stems from the usually fast approach to unity 
of $g(\bm r)$.~\cite{HM06}
Invoking (\ref{eq:comp}), and taking the limit $k \to 0^+$ which discards the value
$\bm k = \bm 0$ to avoid the above diverging term in $\delta(\bm k)$, we have
\begin{equation}
\lim_{~{\bm k}\to 0^+} S_\infty({\bm k})= 
\frac{\chiinf}{\chiinf^{ig}}. 
\label{compr_sk}
\end{equation}
In practice of course, say in a simulation, what is accessible 
is the finite-size $S_L(\bm k)$ and not $S_\infty({\bm k})$,
which in turns allows for the estimation of the bulk 
compressibility through  Eq.~\eqref{compr_sk},
up to finite-size effects. Indeed, the periodic boundaries 
imply that the density of particles, for any microscopic
configuration, is a periodic function. This imposes
a severe restriction on the admissible values of ${\bm k}$, since they must
be commensurate with the periodicity of the system:
\begin{equation}
{\bm k}=\frac{2\pi}{L}(n_{x},n_{y}) 
\label{allowed_values}
\end{equation}
in two dimensions
where $n_{x}$ and $n_{y}$ are any two integers. Hence, $k_{m}\equiv 2 \pi/L$ is the smallest modulus of the allowed non-vanishing vectors. 
As a precursor of the diverging $\delta$ term 
in \eqref{delta_in_k0}, we note that for the finite-size structure factor,
$S_L(0) = N$, and that therefore, the thermodynamic limit of $S_L(\bm k)$
is singular at $\bm k = \bm 0$.
In practice, $\bm k=0$ should be left aside, and an operational way to compute the compressibility
is to consider the large $L$ limit of $S_L(2\pi/L)$.
Our goal is not to discuss what finite-size corrections ensue
(see e.g. refs~\cite{salacuse1,FR13}), but 
to investigate what happens when one mistakenly computes
$S_L(\bm k)$ for $\bm k$ values that are not within
the allowed discrete set \eqref{allowed_values}.
To this end, it is appropriate to revisit the ideal
gas limit, where all quantities are easily derived.
As happened for the density fluctuations in previous
sections, several key feature thereby obtained
do apply to interacting systems as well,
as we shall see below.

\subsection{From the ideal gas...}

\begin{figure}[h!bt] 
\begin{center}
\includegraphics[width=.5\columnwidth,clip=true]{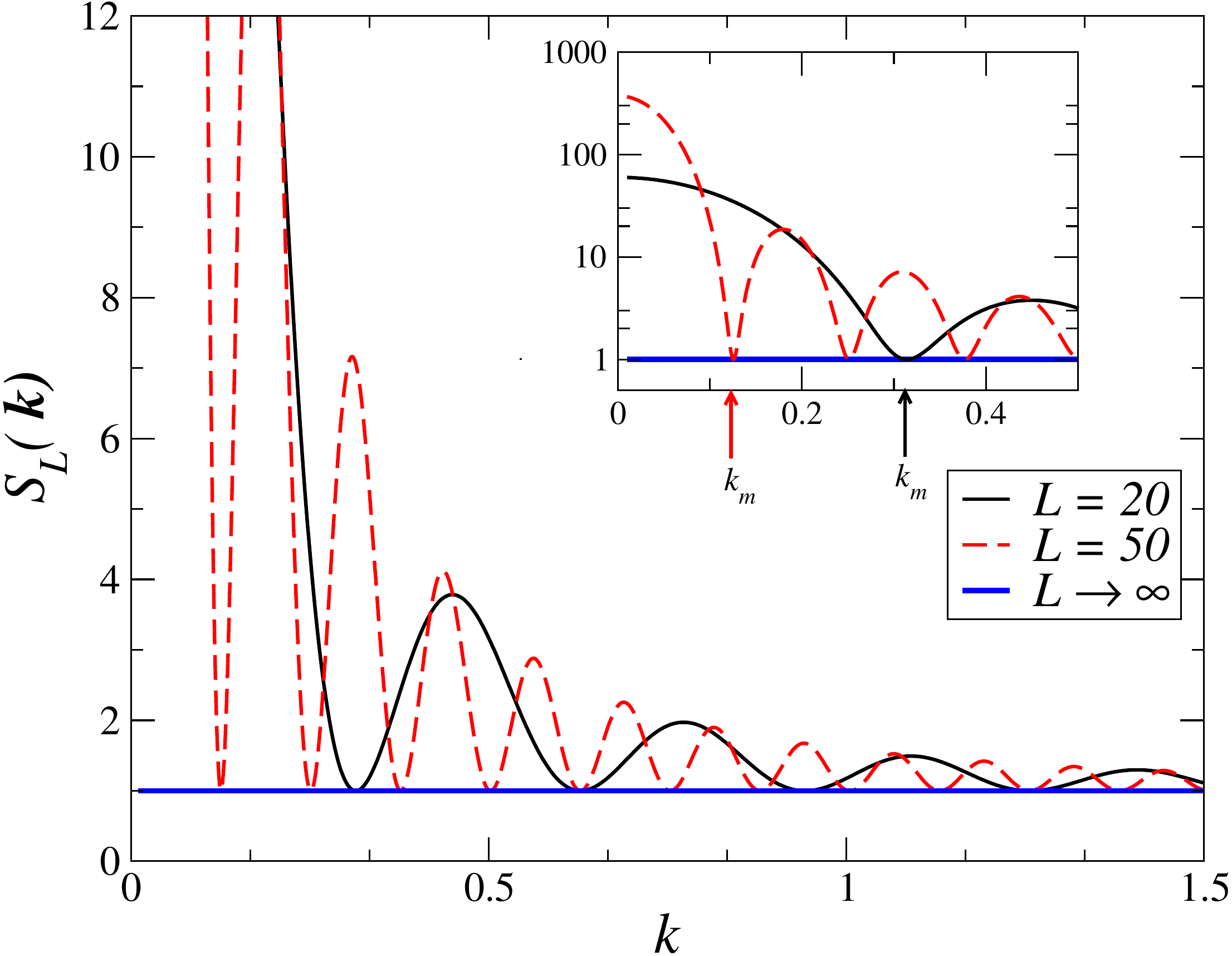}
\caption{\label{fig:struct_ideal} Ideal gas finite size structure factor $S_L({\bm k})$ for different values of $L$.
Here, $S_L$ depends solely on $kL$, so that all curves shown, when plotted in that variable, do collapse onto
the same graph if $N$ is fixed. The goal however is to discuss the 'pollution' stemming from $\delta S_w$
and we therefore choose an arbitrary reference length scale, to illustrate the dependence on $L$.
Here, we have taken $k_y=0$, and $N$ scales as $L^2$, to mimic a constant
density system.
Inset: behavior for small $k\equiv k_{x}$. The
  values of $k_{m}$ (see main text) for the different $L$ are
  shown by the arrows.}
\end{center}
\begin{pspicture}(0,0)(0.5,0.1)
\end{pspicture}
\vspace{-0.8cm}
\end{figure}

For non interacting systems, there is no length scale present
in the model, which results in a constant pair correlation function
and a thermodynamic limit structure factor $S_\infty(\bm k)$ that cannot
depend on $\bm k$. In Eq.~(\ref{ideal_gas_integral}), 
one can set $g_N({\bm r})=1-1/N$ as alluded to in section \ref{ssec:factog}, to cast
the structure factor as a sum of two contributions:
\begin{equation}
S_L({\bm k})=S_\infty({\bm k})+\delta S_{w}({\bm k},L)\label{eq:struct_ideal}
\end{equation}
where $S_\infty({\bm k})\equiv 1$ is the thermodynamic limit value of the structure
factor --indeed structureless-- and $\delta S_{w}({\bm k},L)$ is given by 
\begin{equation}
\delta S_{w}({\bm k},L) \, =\, 
(N-1)\left\vert\frac{4\sin\left(\frac{L}{2} k_{x}\right)\sin\left(\frac{L}{2} k_{y}\right)}{k_{x}k_{y}L^2}\right\vert^{2}. 
\label{wrong_values_sk}
\end{equation}
For the sake of the discussion, 
we restrict to two dimensions, without loss of generality.
The term $\delta S_w$ bears a subscript {\em w} to remind that 
it contributes only when `{\em wrong}' values of $\bm k$ are considered: 
as it should, $\delta S_w(\bm k) = 0$ when $\bm k$ fulfills 
\eqref{allowed_values}, so that the correct result of a
unit structure factor is recovered. In Fig.~\ref{fig:struct_ideal}, the behaviour encoded in Eq.~\eqref{eq:struct_ideal}  is shown for different values of $L$. 
At arbitrary ${\bm k}$ fixed and different from zero
\begin{equation}
\lim_{L\to \infty}\delta S_{w}({\bm k},L)= 0,
\end{equation}
since such a correction must disappear in the thermodynamical limit,
where all the values of ${\bm k}$ are allowed. Let
us also note that at fixed $L$ :
\begin{equation}
\lim_{(k_{x},k_{y})\to 0}\delta S_{w}({\bm k},L) \, =\,
\delta S_{w}({\bm 0},L) \,=\, N-1. 
\label{divergence}
\end{equation}
This shows that the divergence of $S_L(0)=N$ in the thermodynamic limit 
is in some sense due to the non physical contribution $\delta S_w$.
Working on the discrete allowed set (\ref{allowed_values}), and taking 
the limit $k\to 0^+$,
this divergence disappears, and the meaningful compressibility is
obtained.


\subsection{...back to interacting systems}
\label{ssec:btis}

\begin{figure}[h!bt]
\begin{center} 
\includegraphics[width=.6\columnwidth,clip=true]{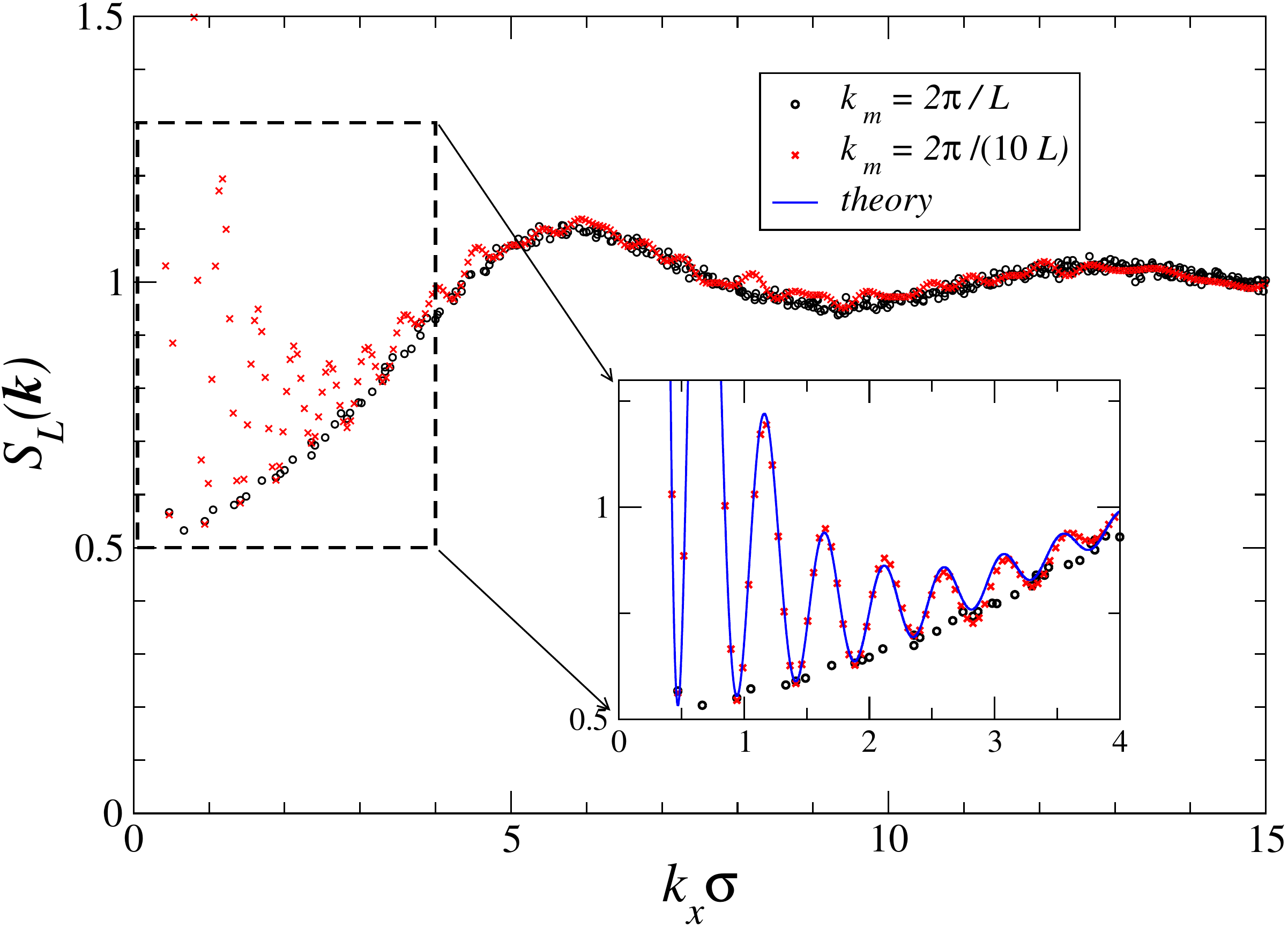}
\caption{\label{fig:str_real} Finite size structure factor $S_L({\bm k})$ in the interacting case, where the function $S_L({\bm k})$
  is shown for a set of allowed values (black circles) and not-allowed
  (red crosses).  
  Here, $k_{y}=0$ and ${\bm k}\equiv (k_{x},0)$. The
  oscillatory behavior, already shown in Fig.~\ref{fig:struct_ideal} for the ideal gas, is
  recovered. Values of the parameters are $N=40$, $\rho^*=0.225$.
  Inset: zoom into the low $k_x$ region. 
  For our purposes, the data shown with circles can be considered as a fair
  approximation to $S_\infty(\bm k)$, while those with crosses in red are 
  for non-allowed Fourier modes, and thus plagued by a non vanishing contribution
  stemming from $\delta S_w$ in \eqref{wrong_values_sk}. The solid line
  in the inset is for the prediction~(\ref{real_system_corrections}). It is in excellent agreement with 
  simulation data, and shows that the apparently scattered crosses at low $k_x$ in the main graph
  are simply a signature of $\delta S_w$.}
\end{center}
\begin{pspicture}(0,0)(0.5,0.1)
\end{pspicture}
\vspace{-0.8cm}
\end{figure}

An interesting feature of relation \eqref{wrong_values_sk} is that it survives
to the ideal gas limitation. It is generically valid for a system of interacting particles
\begin{equation}
 S_L({\bm k})\simeq S_{\infty}({\bm k})+\delta S_{w}({\bm k},L)  
 \label{real_system_corrections}
\end{equation}
up to terms $\mathcal{O}(1/L)$, where $\delta S_{w}({\bm k},L)$ is
the same as in~(\ref{wrong_values_sk}). The sub-leading terms
--discarded here-- have been already calculated in other
studies.~\cite{salacuse1} We focus here on the presence of $\delta
S_w$, stemming from improper account of periodic boundaries, the
consequences of which seem to have been overlooked in some previous
studies.  The derivation of \eqref{real_system_corrections} makes use
of similar arguments as those invoked in section \ref{ssec:factog},
and is reported in
Appendix~\ref{app:structure}. Eq.~\eqref{real_system_corrections} is
well obeyed in molecular dynamics simulations, as shown in
Fig.~\ref{fig:str_real}.  
The oscillations generated by $\delta S_{w}({\bm k},L)$ when $S_L(\bm
k)$ is evaluated for the "wrong $\bm k$ values" are unphysical. Such a
mistake is frequently made by newcomers to the field, but can also be
found in the literature, in different settings. An
  example is given by Refs.~\cite{WG81,WG82}, where two dimensional
harmonic crystals are studied.  The authors work out analytically the
structure factor for different size of a monolayer while working in a
square finite box of length $L$.  All phonons with a wavelength
exceeding $L$ are excluded, so that the range $k<2\pi/L$ is not
accessible, but values larger than this lower cutoff are all
considered as acceptable. This does not comply with the discrete rule
\eqref{allowed_values}. Once corrected for that problem, the computed
finite $L$ structure factors (shown e.g, in Figs. 2 and 3 of
Refs.~\cite{WG81,WG82}), no longer exhibit the spurious oscillations
which are the exact counterpart of those illustrated in
Figs. \ref{fig:struct_ideal} or \ref{fig:str_real}, and are in much
better agreement with the large $L$ analytical predictions than
reported.

\section{Summary}
\label{sec:concl}

We have presented a study of fluctuations and correlations
in a finite-size interacting fluid, where several relevant statistical
physics results can be obtained. Our treatment implicitly addresses
systems with short range interactions, and excludes for instance
Coulombic problems where specific sum rules do apply.\cite{HM06}
Particular emphasis was put 
on a geometry where the confining volume is cubic, and replicated 
through periodic boundary conditions. This is a common 
scheme to minimize surface effects in a numerical simulation,
and thereby emulate bulk phases.~\cite{HM06,FrenkelSmit}
We have shown here how the periodicity of the system
interferes with the fluctuations of the number of particles
in a control volume that exceeds that of the confining box,
through the correlations that are induced between a given particle 
in the central box and its replicated images in neighboring cells.
In addition, several known results in the simpler case where the 
control volume is smaller than the simulation box have been 
rederived in section \ref{sec:inside}, but from an original perspective.
In a second step, we addressed the dual problem 
of computing the static structure factor of the fluid,
but for Fourier modes that do not comply with periodicity. 
Such a procedure, that is met at times, 
yields unphysical results which have been examined. 
In the course of the argumentation, it appeared that the limiting
case of non-interacting systems provided the germane effects under scrutiny and could be 
singled out (factorized) from those stemming from interactions.

\begin{acknowledgments}
We thank David Lacoste, Ladislav \v{S}amaj 
and Juan Antonio White for useful discussions and a careful reading of the manuscript.
\end{acknowledgments}

\bibliographystyle{unsrt}
\bibliography{paper.bib}

\appendix
\section{Definition of correlation functions}
\label{app:A}
From the microscopic density, one defines the single-particle density field:
\begin{equation}
\rho^{(1)}_{N}({\bm r})=\left<\sum_{i=1}^{N}\delta({\bm r}-{\bm r}_{i})\right>\label{micr_dens}
\end{equation}
where the brackets denote equilibrium average.
In the same vein, we introduce the two-particle correlation function
\begin{equation}
\rho^{(2)}_{N}({\bm r,\bm r'})=\left<\sum_{i\neq i'}^{N}\delta({\bm r}-{\bm r}_{i})\delta({\bm r'}-{\bm r}_{i'})\right>.\label{micr_corr}
\end{equation}
Interactions affect pair correlations, that are most conveniently
encoded in the 
pair correlation function
\begin{equation}
g_{N}({\bm r,\bm r'})=\frac{\rho^{(2)}_{N}({\bm r,\bm r'})}{\rho^{(1)}_{N}({\bm r})\rho^{(1)}_{N}({\bm r'})}.\label{pair_corr_func}
\end{equation}
This quantity, measurable from scattering experiments, plays a pivotal
role in the study of simple liquids.~\cite{HM06,C87} In a homogeneous
system like those considered in the present study,
$\rho^{(1)}_{N}({\bm r})$ in independent on $\bm r$ and takes the
value $\rho$ while $\rho^{(2)}_{N}({\bm r,\bm r'})$ and $g_{N}({\bm
  r,\bm r'})$ only depend on the relative position ${\bm r- \bm
  r'}$. It will therefore be denoted $g_N(\bm r)$.  By definition,
$\rho g_N(\bm r)$ is the conditional density of particles at position
$\bm r$, given that a particle sits at the origin. $g_N(\bm r)$ is a
function of density and system size through the number of particles
$N$.  In the thermodynamic limit ($N\to\infty$), it is simply denoted
$g(\bm r)$. This is the quantity shown in the inset of
Fig.~\ref{fig:microzoom}, which as a typical shape for systems
interacting through a steeply repulsive potential at short distances:
for $l<\sigma$, $g(\sigma)\simeq 0$, since particles cannot overlap,
while $g\to 1$ for large $l$, since having a particle at a given point
becomes immaterial for particles at distance $l$ away.

\section{Ideal gas and binomial distribution} 
\label{app:B}

Let us consider an ensemble of $N$ non-interacting particles in a
$d$-dimensional box of length $L$, where by virtue of spatial homogeneity, the probability 
for finding a particle in a sub-region $\mathcal D$ is proportional to its volume $\mathcal{A}(\mathcal{D})$, namely:
\begin{equation}
p_\mathcal{D}=\frac{\mathcal{A}(\mathcal{D})}{L^d}.
\end{equation}
Moreover, the probability for
having $N_\mathcal{D}$ particles in $\mathcal{D}$ is given by the Binomial 
distribution (see e.g.~\cite{SPANISH_AJP}):
\begin{equation}
P_{\mathcal{D}}(N_\mathcal{D})\equiv{N \choose N_\mathcal{D}}p^{N_\mathcal{D}}_{\mathcal{D}}(1-p_{\mathcal{D}})^{N-N_\mathcal{D}},
\label{bern_ideal_gas}
\end{equation}
with mean $E(N_{\mathcal{D}}) = N p_\mathcal{D}$ and 
variance $V(N_\mathcal{D}) = N p_\mathcal{D} (1-p_\mathcal{D})$.
It follows from the definition \eqref{eq:chiL} that 
\begin{equation}
\frac{\chi_{_L}^{ig}(\mathcal{D})}{ \chiinf^{ig} } \, = \,
\frac{\langle N_{\mathcal{D}}^2 \rangle - \langle N_\mathcal{D} \rangle^2}{\langle N_\mathcal{D} \rangle} 
\,=\,
1-p_\mathcal{D} \, = \, 1-\frac{\mathcal{A}(\mathcal{D})}{L^d}.
\label{ideal_gas_compress}
\end{equation}
This is the result quoted in the main text. In the limit where $\mathcal{A}(\mathcal{D})/L^d \to 0$, which we can dub 
the grand canonical condition, the right hand side of
\eqref{ideal_gas_compress} goes to 1. A reformulation of
that result is that for $p_\mathcal{D} \to 0$, that is in the
grand canonical ensemble, the Binomial
law goes to Poisson distribution~\cite{H63}
\begin{equation}
P_\mathcal{D}(n)=\frac{\lambda^n e^{-\lambda}}{n!}
\end{equation}
where $\lambda=  N p_{\mathcal{D}}$ is the mean value.
It is a property of Poissonian variables that the 
mean value and the variance are equal. 

\section{Compressibility and finite size effects}
\label{app:compr}

The quantities appearing in Eq.~\eqref{eq:comp} can be used to define two different objects $\chi_{_L}(\mathcal{D})$ and $\widetilde{\chi}_{g}(\mathcal{D})$ according to:
\begin{equation}
\frac{\chi_{_L}(\mathcal{D})}{\chiinf^{ig}}=\frac{\langle N_{\mathcal{D}}^2 \rangle - \langle N_\mathcal{D} \rangle^2}{\langle N_\mathcal{D} \rangle},\label{app:chil}
\end{equation}
and
\begin{equation}
\frac{\widetilde{\chi}_{g}(\mathcal{D})}{\chiinf^{ig}}= 1+\rho\int_{\mathcal{D}}d{\bm r}\,\left[g({\bm r})-1\right].\label{app:chig}
\end{equation}
When $\mathcal{D}$ is large in an even much larger (say infinite) system, $\widetilde{\chi}_{g}(\mathcal{D})$ and $\chi_{_L}(\mathcal{D})$ coincide with the bulk compressibility $\chi_{\infty}$. In general, however, these two quantities differ as we now illustrate.

\begin{figure}[h!bt] 
\begin{center}
\includegraphics[width=.5\columnwidth,clip=true]{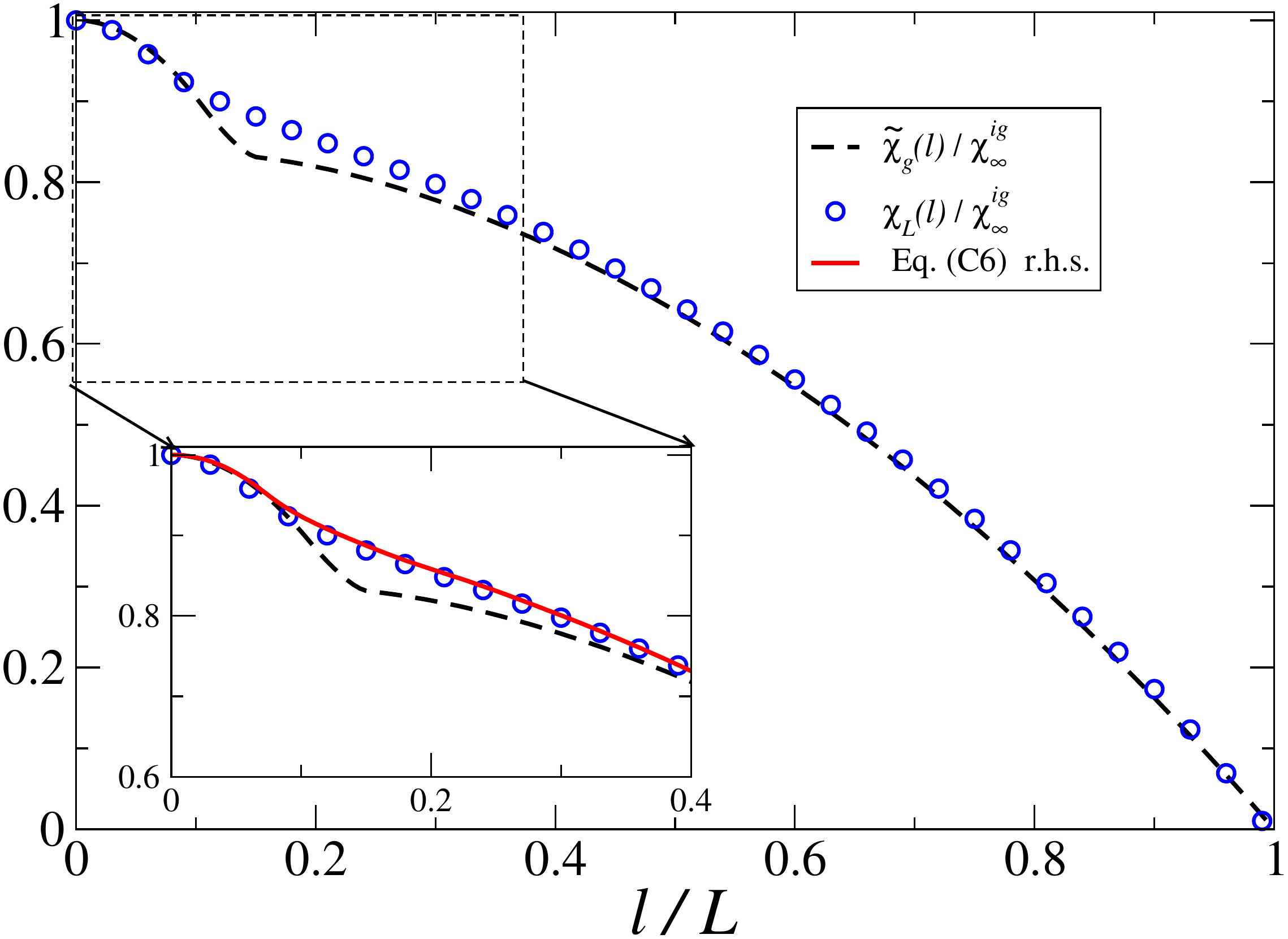}
\caption{Numerical evaluation of $\chi_{_L}(\mathcal{D})$ and
  $\widetilde{\chi}_{g} (\mathcal{D})$ in the interacting
  model, with $\rho^{*}\simeq 0.07$. Inset: zoom into the microscopic region (small $l$) and verification of Eq.~\eqref{eq:real_correction}. \label{fig:gr_corr}}
\end{center}
\end{figure}

In order to see the connection between Eqs.~\eqref{app:chil}
and~\eqref{app:chig} it is necessary to start from the definition of
particle fluctuations via the density field~\eqref{micr_dens}, namely:
\begin{equation}
\left<\left( N_{\mathcal{D}}-\left<N_{\mathcal{D}}\right>\right)^{2}\right>=\left<\int_{\mathcal{D}^{2}}d{\bm r}_{1}d{\bm r}_{2}\,\left(\rho^{(1)}_{N}({\bm r_{1}})-\rho\right) \left(\rho^{(1)}_{N}({\bm r_{2}})-\rho\right)\right>.
\end{equation}
By using the definitions in Eqs.~(\ref{micr_dens}-\ref{pair_corr_func}) this expression
can be rewritten in terms of the radial distribution function:
\begin{equation}
\left<N^{2}_{\mathcal{D}}\right>-\left<N_{\mathcal{D}}\right>^{2}=\rho^{2}\int_{\mathcal{D}^2}d{\bm r}_{1}d{\bm r}_{2}\,\left[g_{N}({\bm r}_{1}-{\bm r}_{2})-1\right]+\rho \mathcal{A}(\mathcal{D}).\label{fluct_corr}
\end{equation}
To take advantage of the fact that $g_{N}$ only depends
on relative position in a homogeneous system, one usually changes
variables from $\{{\bm r}_{1},{\bm r}_{2}\}$ to $\{{\bm r}\equiv{\bm r}_{1}-{\bm r}_{2},{\bm r}_{2}\}$, but attention should be paid to the domain of integration, that becomes distorted. An expedient way of proceeding is to consider the new domain $\widetilde{D}\equiv\{{\bm
  r}_{1}-{\bm r}_{2}\vert {\bm r}_{1}\in \mathcal{D},{\bm r}_{2}\in
\mathcal{D}\}$, and insert the identity
\begin{equation}
1\equiv \int_{\widetilde{\mathcal{D}}} d {\bm r}\,\delta({\bm r}-{\bm r}_{1}+{\bm r}_{2}  ),\label{identity}
\end{equation}
into Eq.~\eqref{fluct_corr}. We stress that Eq.~\eqref{identity} is valid only if ${\bm r}_{1}$ and ${\bm
    r}_{2}$ belong to the set $\mathcal{D}$, which is our case,
  otherwise the left hand term must be substituted with a Heaviside step
  function. We thus get
\begin{equation}
\frac{\left<N^{2}_{\mathcal{D}}\right>-\left<N_{\mathcal{D}}\right>^{2}}{\left<N_{\mathcal{D}}\right>}=1+\rho\int_{\widetilde{\mathcal{D}}}d{\bm r}\,w_{\mathcal{D}}({\bm r})\left[g_{N}({\bm r})-1\right],\label{eq:real_correction}
\end{equation}
where
\begin{equation}
w_{\mathcal{D}}({\bm r})=\frac{\int_{\mathcal{D}^2}d{\bm r}_{1}d{\bm r}_{2}\,\delta({\bm r}-\vert{\bm r}_{1}-{\bm r}_{2}\vert  )}{\mathcal{A}(\mathcal{D})} \label{definition_corr}
\end{equation}
is a dimensionless weight function. This is exactly the expression we
were looking for, connecting fluctuations to a proper integral of the
radial distribution function for arbitrary domain $\mathcal{D}$, no
matters how small. Remarkably, Eq.~\eqref{eq:real_correction} yields
the right hand side of the ordinary formula (\ref{eq:comp}) in two
limiting cases only:
\begin{itemize}
\item when $\mathcal{A}(\mathcal{D})$ is small with particles having an hard-core like repulsion, since $g({\bm r})\simeq 0 $ and $\int_{\widetilde{\mathcal{D}}}d{\bm r}\,w_{\mathcal{D}}({\bm r})=\int_{\mathcal{D}}d{\bm r}=\mathcal{A}(\mathcal{D})$.
\item when $\mathcal{A}(\mathcal{D})$ is large, since $w_{\mathcal{D}}({\bm r})\to 1$ when  $\mathcal{D}\to \mathds{R}^{d}$, and the integral over $\widetilde{\mathcal{D}}$ coincides with that one over the whole space, and with that over $\mathcal{D}$. 
\end{itemize}
From the limits above, it appears that the maximum deviation between $\chi_{_L}(\mathcal{D})$ and $\widetilde{\chi}_{g}(\mathcal{D})$ appears in some intermediate regime, as confirmed in Fig.~\ref{fig:gr_corr}. For instance, in the case of the squared shape region described in the main article (length $l$ and dimension $d$) a straightforward calculation from~\eqref{definition_corr}, gives
\begin{equation}
w_{l}({\bm r})=\prod_{i}^{d}\left(1-\frac{\vert x_{i}\vert}{l}\right),\qquad \textrm{where } {\bm r}\equiv\{x_{1}, x_{2}, \dots, x_{d}\}.\label{gr_corr_square}
\end{equation}
In the inset of Fig.~\ref{fig:gr_corr}, one can observe that Eq.~\eqref{eq:real_correction}, is well obeyed.

\section{A result on the fluctuations of correlated variables} 
\label{sec:app}
Let us
consider a box which contains exactly $N$ homogeneously distributed particles.
We define two non overlapping regions,
$\mathcal{D}_{1}$ and $\mathcal{D}_{2}$, and we are interested
in the statistics of the following quantity:
\begin{equation}
N_{t}=\alpha N_{1}+\beta N_{2}, \label{weighted_sum}
\end{equation} 
where $N_{i}$ is the number of particles
contained in the region $\mathcal{D}_{i}$. Eq.~\eqref{weighted_sum} is the linear combination
of two correlated random variables (for instance, when $N_{1}=N$,
$N_{2}$ necessarily vanishes). The calculation of the mean values,
denoted again with the symbol $E$, simply follows from the homogeneity assumption
\begin{equation}
E(N_{t})=\alpha E(N_{1})+\beta E(N_{2})=\alpha N p_1+ \beta N p_2
\end{equation}
with $p_i\propto \AD{i}$, the volume of region 
${\cal D}_i$. When it comes to computing the variance $V(N_{t})$, a result known as the law of total
variance~\cite{W05} turns useful. It reads:
\begin{equation}
V(N_{t})=V(E(N_{t}\vert N_{3}))+E(V(N_{t}\vert N_{3})), \label{app:equality}
\end{equation}
where $N_{3}$ is any arbitrary variable, used for
conditioning. Loosely speaking, Eq.~\eqref{app:equality} states that
the total variance is the mean of the conditioned variance plus the
variance of the conditioned mean. More precisely, $E(N_t\vert N_3)$
(resp $V(N_t\vert N_3)$) signifies that $N_3$ being fixed, one computes the corresponding mean value (respectively variance) of $N_{t}$.
In the present situation, it is convenient to choose $N_{3}=N_1+N_2$, the number of particles in $\mathcal{D}_3=\mathcal{D}_1\cup\mathcal{D}_2$. Then,
\begin{equation}
E(N_t\vert N_3)=\alpha E(N_1\vert N_3)+ \beta E(N_2\vert N_3)=\alpha N_3 \frac{p_1}{p_1+p_2}+\beta N_3\frac{p_2}{p_1+p_2},
\end{equation}
as follows from the homogeneity assumption. Indeed, when $N_3$
particles exactly lie in $\mathcal{D}_{3}$, a fraction $\frac{\AD{i}}{\AD{1}+\AD{2}}$ of them lies in $\mathcal{D}_{i}$ on average.
For the (unconditioned) variance, we have
\begin{equation}
V(E(N_t\vert N_3))=\left(\frac{\alpha p_1+\beta p_2}{p_1+p_2}\right)^{2}V(N_3).
\end{equation}
We turn to the second contribution in~Eq.~\eqref{app:equality}
\begin{eqnarray}
V(N_t\vert N_3)& =& V(\alpha N_1+\beta (N_3-N_1)\vert N_3)\nonumber \\
&=& V( (\alpha-\beta) N_{1}+\beta N_3\vert N_3)\nonumber \\
&=& V((\alpha-\beta)N_1\vert N_3),
\end{eqnarray}
since the variance is not affected by a shift of the variable. Thus
\begin{equation}
V(N_t\vert N_3) \, = \, (\alpha-\beta)^2 \, V(N_1\vert N_3).
\end{equation}
Gathering results, one gets:
\begin{equation}
V(N_{t}) \, = \, \left(\frac{\alpha p_1+\beta p_2}{p_1+p_2}\right)^{2}V(N_{3}) \,+ \, (\alpha-\beta)^{2} \,E(V(N_{1}| N_{3})) \label{Variance}
\end{equation}
Note when $\alpha=\beta = 1$, $N_t=N_1+N_2=N_3$ and we recover the tautological
expression $V(N_{t})=V(N_{3})$.

\section{Structure factor and finite size effects}
\label{app:structure}
In this section, we derive Eq.~\eqref{real_system_corrections} for the structure factor, as considered in
Sec.~\ref{sec:structure}. In this case, there are two different sources
of finite-size effects. Indeed, apart from the finite number of
particles considered, the finite length of the box imposes a strong
constraint on the allowed set of wave vectors. In order to derive the
final result, both restrictions must be taken into account. For simplicity, we study the two dimensional case but what follows can be easily generalized to other dimensions.

We consider two different boxes of lengths $R$ and $L$ with $L<R$ and
their finite size structure factors  $S_{R}({\bm k})$ and  $S_L({\bm k})$. 
The \emph{allowed} values of ${\bm k}$ with respect to the box
$R$ belong to the set $U_{R}=\{\frac{2\pi}{R}{\bm n}\}$ where ${\bm n}$ is any vector with integer coordinates,
see (\ref{allowed_values}). 
In the thermodynamical limit, the structure factor sought for is recovered:
\begin{equation}
\lim_{R\to +\infty} S_{R}({\bm k}) = S_{\infty}({\bm k})
\end{equation}
and in the same limit
$U_{R}\to \mathds{R}^{2}
$ while all the values of ${\bm k}$ are allowed. By definition, the
structure factor is connected to the pair distribution function. From Eq.~(\ref{ideal_gas_integral}), we have that:
\begin{eqnarray}
S_{R}({\bm k}) &=& 1+\frac{N}{R^{2}}\int_{\widetilde{B}_{R}} d{\bm r}\, w_{R}({\bm r}) g_{N,R}({\bm r}) \exp\left[-i{\bm k} \cdot {\bm r}\right]=\\
& = & 1+\frac{N}{R^{2}}\int_{\widetilde{B}_{R}} d{\bm r}\, g_{N,R}({\bm r}) \exp\left[-i{\bm k} \cdot {\bm r}\right]+\mathcal{O}(R^{-1})\\
\end{eqnarray}
where attention should be paid to the fact that ${\bm r}\equiv {\bm  r}_{2}-{\bm r}_{1}$ denotes the relative position of the particles,
so that if the available volume for each position ${\bm r}_1$ and
${\bm r}_2$ is a square, then $\widetilde{B}_R$ is a
parallelogram. For this reason the same weight function $w_{R}({\bm r})$, considered in
Sect.~\ref{app:compr}, appears. However, this point is not crucial for
what follows.

Let us now consider the region $B_{L}$, square-shaped of length $L$,
with $L<R$. If for some integer $n$, one has that
\begin{equation}
 R=nL, \label{commens}
\end{equation}
then clearly the values allowed for $B_{L}$ are a subset of
$U_{R}$, but this is not true in general.
We next evaluate the following difference
\begin{equation}
S_L({\bm k})-S_{R}({\bm k})=  I_{1}+I_{2}+\mathcal{O}(R^{-1})+\mathcal{O}(L^{-1})  \label{app:separation}
\end{equation}
where
\begin{eqnarray}
I_{1}&=&\rho \int_{\widetilde{B}_{L}} d{\bm r}\, e^{-i {\bm k}\cdot{\bm r}} \left[ g_{N,L}({\bm r})-g_{N,R}({ \bm r})\right]\\ 
I_{2}&=&-\rho \int_{\widetilde{B}_{R}-\widetilde{B}_{L}} d{\bm r}\,  e^{-i {\bm k}\cdot{\bm r}}g_{N,R}({\bm r}) 
\end{eqnarray}
We stress that Eq.~\eqref{app:separation} is considered only for the values of $\bm k$ that belongs to $U_{R}$.
The two terms are now analyzed separately. 
The first one gives a $1/N$ contribution. In order to show that, we recall the
expansion used in Eq.~\eqref{eq:gr_exp}:
\begin{equation}
g_{N,R}({\bm r})=g({\bm r})+\delta g_{R}
\end{equation}
where $\delta g_{R}\sim \mathcal{O}\left(1/N\right)=\mathcal{O}\left(1/R^{2}\right)$, together with the corresponding equation for $g_{N,L}({\bm r})$. One has that
\begin{equation}
I_{1}=\rho \int_{\widetilde B_{R}} d{\bm r} e^{-i {\bm k}\cdot{\bm r}} \left[ \delta g_{L}-\delta g_{R}\right].
\end{equation}
Then in the thermodynamical limit
\begin{equation}
\lim_{R\to +\infty} I_{1}=  \int_{\widetilde B_{R}} d{\bm r}\, \delta g_{L}\, e^{-i {\bm k}\cdot{\bm r}}\sim \mathcal{O}\left(L^{-2}\right).\label{app:partial1}
\end{equation}

Regarding the second contribution, under the condition $R\gg L\gg 1$ and since the integrand runs for distances larger than $L$, we can assume $ g_{N,L}({\bm r}) \simeq 1$, apart from a term $\mathcal{O}\left(L^{-2}\right)$, and then
\begin{equation}
I_{2}\simeq-\rho \int_{\widetilde{B}_{R}-\widetilde{B}_{L}} d{\bm r}\,  e^{-i {\bm k}\cdot{\bm r}} = \rho \int_{\widetilde{B}_{L}} d{\bm r}\,  e^{-i {\bm k}\cdot{\bm r}}=\delta S_{w}({\bm k},L),\label{app:partial2}
\end{equation}
since for the set $U_{R}$, the integral over $\widetilde{B}_{R}$ vanishes. As
already defined in Sec.~\ref{sec:structure}, the last integral in
Eq.~\eqref{app:partial2} leads to the ideal gas correction that we
rewrite here for completeness:
\begin{equation}
  \delta S_{w}({\bm k},L)= \frac{(N-1)}{L^{4}}\left\vert\frac{4\sin\left(\frac{L}{2} k_{x}\right)\sin\left(\frac{L}{2} k_{y}\right)}{k_{x}k_{y}}\right\vert^{2}. \label{app:wrong_values_sk}
\end{equation}
As appears from Eq.~\eqref{app:wrong_values_sk}, $I_{2}$
contains the contribution of the wrong ${\bm k}$: it vanishes as soon as $B
_{L}$ is commensurate with $B_{R}$, since $U_{L}$ is contained in $U_{R}$. 
Inserting Eq.~(\ref{app:partial1}) and Eq.~(\ref{app:partial2}) into
Eq.~\eqref{app:separation}, in the thermodynamical limit where $R\to
\infty$, one obtains the final result
\begin{equation}
  S_L({\bm k})= S_{\infty}({\bm k})+\delta S_{w}({\bm k},L) +\mathcal{O}\left(L^{-1}\right).\label{app:real_system_corrections}
\end{equation}
This is the expression, valid for all ${\bm k}$, which is considered in section \ref{ssec:btis}.

\end{document}